\documentclass[12pt]{iopart}
\usepackage{graphicx}
\usepackage{ulem}
\usepackage{color}
\usepackage{iopams}
\usepackage{setstack}
\usepackage{mathabx}

\begin{document}

\title{Recognition Dynamics in the Brain under the Free Energy Principle}

\author{Chang Sub Kim}
\address{Department of Physics,
Chonnam National University,
Gwangju 61186, Republic of Korea}
\ead{cskim@jnu.ac.kr}

\begin{abstract}
We formulate the computational processes of perception in the framework of the principle of least action by postulating the theoretical action as a time integral of the free energy in the brain sciences.
The free energy principle is accordingly rephrased as that for autopoietic grounds all viable organisms attempt to minimize the sensory uncertainty about the unpredictable environment over a temporal horizon.
By varying the informational action, we derive the brain's recognition dynamics (RD) which conducts Bayesian filtering of the external causes from noisy sensory inputs.
Consequently, we effectively cast the gradient-descent scheme of minimizing the free energy into Hamiltonian mechanics by addressing only positions and momenta of the organisms' representations of the causal environment.
To manifest the utility of our theory, we show how the RD may be implemented in a neuronally based biophysical model at a single-cell level and subsequently in a coarse-grained, hierarchical architecture of the brain.
We also present formal solutions to the RD for a model brain in linear regime and analyze the perceptual trajectories around attractors in neural state space.
\end{abstract}


\maketitle

\section{Introduction}
\label{Introduction}
\noindent

The quest for a universal principle that may explain the cognitive and behavioral operation of the brain is of great scientific interest at present.
The apparent difficulty in addressing the quest is the gap between the information processing and the biophysics that governs neurophysiology in the brain.
However, it is evident that the matter, which is the ground-stuff upon which the brain functions emerge, comprises neurons obeying the laws guided by physics principles.
Thus, any biological principles that attempt to explain the brain's large-scale workings must cope with our accepted physical reality \cite{Erwin}.
It appears that on the current approaches still prevails the classical,  effective epistemology of regarding perceptions as constructing hypotheses which may hit upon truth by producing symbolic structures matching physical reality \cite{Helmholtz1909,Gregory1980,Dayan:1995}.

One influential candidate at present that seeks for such a rubric in neuroscience is the free energy principle (FEP) \cite{Friston2009Trend,Friston2010,Friston:JRSoc2013}.
For a technical appraisal of the FEP, we refer to \cite{Kim2017} where the theoretical assumptions and the mathematical structure involved in the FEP are reviewed in great detail.
We have noticed  \cite{Ramstead:2017} suggesting `variational neuroethology' which explains, integrating the FEP with evolutionary systems theory,  how living systems appear to resist the second law of thermodynamics.
To state compactly, the FEP offers that all viable organisms perceive and act on the external world by instantiating a probabilistic causal model embodied in their brain in a way to ensure their adaptive fitness or autopoiesis \cite{Maturana:1980}.
The biological mechanism that endows the organism's brain with the operation is theoretically framed into an information-theoretic measure, which we call  `informational free energy (IFE)'.
According to the FEP, a living system tries to minimize the sensory surprisal when it faces an external cause that perturbs its spontaneous equilibrium within its physiological boundary by pursuing perceptive as well as active inferences.
However, the brain does not preside over instreaming sensory distribution; accordingly, the brain cannot directly minimize the sensory surprisal but, instead, minimizes its upper bound, the IFE.
The probabilistic rationale of the FEP argues that the brain's representations of the uncertain environment are the sufficient statistics, e.g., means or variances, of a probability density encoded in the brain.
The variational parameters are supposed to be encoded as physical variables in the brain.
The brain statistically infers the external causes of sensory input by Bayesian filtering, using its internal top-down model about predicting, or \textit{generating}, the sensory data.
Filtering is a probabilistic approach to determining the external states from noisy measurements of sensory data \cite{Jazwinski:1970}.
There is growing experimental support for the brain's maintaining internal models of the environment to predict sensory inputs and to prepare actions, see for instance \cite{Berkes2011}.
The computational operation of  the abductive inference is subserved by the brain variables and
the resulting  perceptual mechanics is termed as the `recognition dynamics (RD)'.

Although the suggestion of the FEP has been promising to account for the brain's inferring mechanism of and acting upon sensory causes, we find certain theoretical subtleties in the conventional formulation:
\begin{itemize}
\item [] First, the FEP minimizes the IFE at each point in time for successive sensory inputs \cite{Frsiton:BiolCybern2010}. However, precisely the objective function to be minimized is the continuously accumulated IFE over a finite time.\footnote{According to the FEP, the updating or learning of the generative model takes places in the brain on a longer time scale than that associated with perceptual inference. To derive the RD of the slow variables for synaptic efficacy and gain, the time-integral of the IFE is taken as an objective function; however, again the gradient descent method is executed in a pointwise way in time \cite{Friston:2007}.} The minimization must be performed concerning trajectories over a temporal horizon across which an organism encounters with atypical events to its natural habitat and biology.
\item[] Second, the FEP employs the gradient-descent method in practically executing minimization of the IFE \cite{Friston:GF2010}, which is widely used in machine learning theory to solve engineering optimization problems efficiently. The engaged scheme allows formulation to find  heuristically optimal solutions in the FE landscape, but it not derived from a scientific principle.
\item[] Third, the FEP introduces the notion of the `generalized coordinates' of an infinite number of the so-called `generalized motions' to account for the dynamical nature of the environment \cite{Friston:VF2008}. The ensuing theoretical construct is a generalization of the standard Newtonian mechanics.\footnote{The mechanical state of a particle is specified only by position and velocity in the Newtonian mechanics, and no physical observables are assigned to the dynamical orders beyond the second-order.  In some literature \cite{Sprott1997}, the concept of `jerk' is assigned to the third-order time-derivative of position as a physical reality. From the mathematical perspective, such a generalization is not forbidden. However, not only higher-orders are difficult to measure, but more seriously it raises the question of what the corresponding cause to jerk as the force to acceleration.  And, the same impasse in all next orders.} With the hired theory, however, it is obscure to decide the number of independent dynamical variables for a complete description. In practice, typically dynamic truncation is made at a finite embedding order by assuming that the precision of random fluctuations on higher orders of motion disappears very quickly.
\item[] Fourth, the FEP introduces the hydrodynamics-like concepts of the `path of a mode (motion of expectation)' and the `mode of a path (expected motion)' by distinguishing the dynamic update from the temporal update of a time-dependent state \cite{Friston:PLoS2008}. Because the distinction is essential to ensure an equilibrium solution to the RD in employing the dynamical generative models, further theoretical exploration seems worthwhile.
\item[] Fifth, the FEP considers the states of the environment `hidden' because what the brain faces is only a probabilistic sensory mapping. Subsequently, a distinction is made between the hidden-state representations, responsible for intra-level dynamics, and causal-state representations,  responsible for inter-level dynamics, in the hierarchical brain \cite{Friston:Physiology2006}. Such a distinction must emerge as neuronal dynamics in the brain on different timescales. Accordingly, a biophysically grounded formulation that supports the top-down idea is required.
\end{itemize}

In this paper, we present a mechanical formulation of the RD in the brain in the framework of Hamilton's principle of least action \cite{Landau}.
Motivated by the aforementioned theoretical observations, we try to resolve some of the technical complexities in the FEP framework.
To be specific, the goal is to recast the gradient-descent strategy of minimizing the IFE into the mathematical framework that appeals to the normative physics rules.
We do this by hypothesizing the IFE as a Lagrangian of the brain which enters the theoretical action, being the fundamental objective function to be minimized in continuous time under the principle of least action.
Consequently, we reformulate the RD regarding only the canonical, physical realities to eschew the generalized coordinates of infinitely recursive time-derivatives of the continuous states of the organism's environment and brain.
In the canonical description, the dynamical state of a system is specified only by positions and their first-order derivatives.

In this work, supported by the present day evidence \cite{Markov:2014,Michalareas:2016}, we admit the bi-directional facet in informational flow in the brain.
The environment begets sensory data at the brain-environment interface such as sensory receptors or interoceptors within an organism.
The incited electro-opto-chemical interaction in sensory neurons must transduce forward in the anatomical structure of the brain.
Whereas complying with the idea of perception as constructing hypotheses, there must be backward pathway as well in information processing in the functional hierarchy of the brain.
To understand how such bidirectional functional architecture is emergent from the electrophysiology of biophysics and anatomical organization of the brain is a forefront research interest (see, for instance, \cite{MarkovKennedy2013} and references therein).
We shall consider a simple model that effectively incorporates the functional hierarchy while focusing our attention on the brain's perceptual mechanics of inferring of the external world, given sensory data.
The problem of learning of the environment via updating the internal model of the world and of active inference of changing sensory input via action on the external world, see for instance \cite{Friston:PLoS2009},  is deferred for an upcoming paper.

 Here, we shall outline how in this work we cast Bayesian filtering in the FEP using a variational principle of least action and how we articulate the minimization of the sensory uncertainty in terms of an associated Lagrangian and Hamiltonian. Furthermore, given a particular form for the differential equations, afforded by computational neuroscience, one can see relatively easily how neuronal dynamics could implement the Bayesian filtering:
(i) According to the FEP, the brain represents the environmental features statistically efficiently,  using the sufficient statistics $\mu$.  We assume that $\mu$ stands for the basic computational unit of the neural attributes of perception in the brain. Such a constituent is considered a `perceptual particle' which may be a single neuron or physically coarse-grained multiple neurons forming a \textit{small} particle;
(ii) We postulate that the Laplace-encoded IFE in the FEP, denoted as $F$ (Sec. \ref{IFE}), serves as an effective, informational Lagrangian (IL) of the brain, written as ${\cal L}$. Accordingly, the informational action (IA), which we denote by ${\cal S}$, is defined to be time-integral of the approximate IFE (Sec.~\ref{LagFor1});
(iii) Then, conforming to the Hamilton principle of least action, the equations of motion of the perceptual particles are derived mathematically by varying the IA with respect to both $\mu$ and $\dot\mu$.
The resulting Lagrange equations constitute the perceptual mechanics, i.e., the RD of the brain's inferring of the external causes of the sensory stimuli (Sec.~\ref {LagFor1});
(iv) In turn, we obtain the brain's informational Hamiltonian ${\cal H}$ from the Lagrangian via a Legendre transformation. Consequently, we derive a set of coupled, first-order differential equations for $\mu$ and its conjugate $p_\mu$, which are equivalent to the perceptual mechanics derived from the Lagrange formalism.
The resulting perceptual mechanics is our derived RD in the brain. Accordingly, the brain performs the RD in the state space spanned by the position $\mu$ and momentum $p_\mu$ variables of the constituting neural particles. (Sec.~\ref{HamFor1});
(v) We adopt the Hodgkin-Huxley (H-H) neurons as biophysical neural correlates which form the basic perceptual units in the brain. We first derive the RD of sensory perception at a single-neuron level where the membrane potential, ionic transport, and gating are the relevant physical attributes.
Subsequently, we scale up the cellular formulation to furnish a functional hierarchical-architecture of the brain.
On this coarse-grained scale, the perceptual states are the averaged properties of many interacting neurons.
We simplify the hierarchical picture with two averaged, activation and connection variables, mediating the intra- and inter-level dynamics, respectively.
According to our formulation of the hierarchical RD in the brain,
as sensory perturbation comes in at the lowest level, i.e., sensory interface, the brain carries out the RD in its functional network and finds an optimal trajectory which minimizes the IA.

To summarize, we have adopted the IFE as an informational Lagrangian of the brain and subsequently employed the principle of least action to construct the Hamiltonian mechanics of cognition. In doing so, only positions and momenta of the neural particles have been addressed as dynamical variables; positions and momenta are the metaphorical terms for the perceptual states and the brain's prediction errors, respectively.
We do not distinguish the causal and hidden states, both of which must emerge as biophysical neuronal activities on different timescales.
The resulting RD is statistically deterministic, arising from unpredictable motions of the environmental states and noisy sensory mapping.
Furthermore, the derived RD describes not only the temporal development of the brain variables but also the prediction errors.
The rate of the prediction errors is not incorporated in the conventional formulation of the FEP.
The successful solutions of the RD are stable equilibrium trajectories in the neural state space, specifying the tightest upper bound of the sensory uncertainty, conforming to the rephrased FEP.
Our formulation allows solutions in an analytical form in linear regimes near fixed points, expanded in terms of the eigenvectors of the Jacobian and, thus, provides with tractability of real-time analysis.
We hope that our theory will motivate further investigations of some model brains with numerical simulations and also of the active inference and learning problems.

This paper is organized as follows.
We first recapitulate the FEP in Sec.~\ref{Sec-FEP} to support our motivation for casting the gradient descent scheme into the standard mechanical formulation.
In the followed Sec.~\ref{IAP} we present the RD reformulated in the Lagrangian and Hamiltonian formalisms.
Then,  in Sec.~\ref{Biophysics} biophysical implementations of our theory at the cellular level and in the scaled-up hierarchical brain are formulated, where nonlinear as well as linear dynamical analyses are carried out.
Finally, a discussion is provided in Sec.~\ref{Conclusion}.

\section{The free energy principle}
\label{Sec-FEP}
\indent
To unveil our motivation for this paper, we shall compactly digest here the IFE and the FEP, that are currently exercised in the brain sciences.
The RD is an organism's organization of executing minimization of the IFE in the brain under the FEP.
In practice, there are various ways of IFE-minimizing schemes, for instance, variational message passing and belief propagation, which do not lend themselves to treatment regarding generalized coordinates of motion.
Our treatment in this paper is more relevant to Bayesian filtering and predictive coding schemes that have become a popular metaphor for message passing in the brain.  Filtering is the problem of determining the state of a system from noisy measurements \cite{Jazwinski:1970}.
For a detailed technical appraisal of the FEP, we refer to \cite{Kim2017} from which we borrow the mathematical notations.

\subsection{The informational free energy}
\label{IFE}
A living organism occupies a finite space and time in the unbounded, changing world while interacting with the rest of the world, comprising its environment.
The states of the environment are denoted as $\vartheta$ collectively, which are `hidden' from the organism's perspective.
The signals from the environment are registered biophysically at the organism's sensory interface as sensory data $\varphi$.

The organism's brain faces uncertainty when it tries to predict the sensory inputs, the amount of which is quantified as the \textit{sensory uncertainty} $H$.
The sensory uncertainty is defined to be an average of the \textit{self-information}, $-\ln p(\varphi)$, over the probability density $p(\varphi)$ encoded at the interface,
\begin{equation}\label{sensory-uncertainty}
H\equiv \int d\varphi \left\{-\ln p(\varphi)\right\}p(\varphi).
\end{equation}
The self-information, which is also termed as the sensory `surprise' or `surprisal' in information theory,  quantifies the survival tendency of living organisms in the unpredictable environment.
Assuming that the sensory density describes an ergodic ensemble of sensory streaming\footnote{This ergodicity assumption is an essential ingredient of the FEP, which hypothesizes that the ensemble average of the surprisal is equal to the time-average of that, regarding the surprisal as a statistical, dynamical quantity.}, one may convert the sensory uncertainty into a time-average as
\[
\int d\varphi \left\{-\ln p(\varphi)\right\}p(\varphi) = \frac{1}{T}\int_0^T dt\left\{-\ln p(\varphi(t))\right\},
\]
where $T$ is the temporal window over which an environmental event takes places, i.e., a temporal horizon.
Here, one may manipulate the right-hand side (RHS) of the preceding equation by adding a Kullback-Leibler divergence to the integrand to get
\[ - \ln p(\varphi)  + \int d\vartheta q(\vartheta)\ln\frac{q(\vartheta)}{p(\vartheta |\varphi)}
\rightarrow\int d\vartheta q(\vartheta)\ln\frac{q(\vartheta)}{p(\vartheta,\varphi)}.\]
The outcome brings about the mathematical definition of the IFE,
\begin{equation}
{\cal F}[q(\vartheta),p(\vartheta,\varphi)] \equiv \int d\vartheta q(\vartheta)\ln\frac{q(\vartheta)}{p(\vartheta,\varphi)}, \label{FE-functional}
\end{equation}
which is expressed as a \textit{functional} of the two probability densities, $q(\vartheta)$ and $p(\vartheta,\varphi)$;
where $q(\vartheta)$ and $p(\vartheta,\varphi)$ are termed the recognition density (R-density) and the generative density (G-density), respectively.
The R-density is the organism's probabilistic representation of the external world, which the organism's brain uses in approximately inferring the causes $\vartheta$ of inputs $\varphi$.
The G-density, a joint probability between $\vartheta$ and $\varphi$, underlies the top-down model about how the sensory data are biophysically generated by interaction between the brain and the environment.
By construction, the surprisal is smaller than the IFE by the added positive-amount, accordingly the sensory uncertainty is bounded from above in accordance with
\begin{equation}\label{Math-FEP1}
\int dt\left[-\ln p(\varphi)\right] \le \int dt{\cal F}[q(\vartheta),p(\vartheta,\varphi)].
\end{equation}
Note that the sensory uncertainty on the left-hand side (LHS) of Eq.~(\ref{Math-FEP1}) specifies the accumulated surprisal  over a temporal horizon involved in an environmental event.

Equation~(\ref{Math-FEP1}) constitutes a mathematical statement of the FEP:
``All viable organisms try to avoid being placed in an atypical situation in their environmental habitats for existence by minimizing the sensory uncertainty.
However, organisms do not possess direct control over the sensory distribution $p(\varphi)$; accordingly they, instead, are to minimize the upper bound  of Eq.~(\ref{Math-FEP1}), $\int dt{\cal F}$ as a proxy for the sensory uncertainty.''
The brain conducts the minimization probabilistically by updating the R-density to approximate the posterior density $p(\vartheta |\varphi)$, namely carrying out Bayesian inference of the causes $\vartheta$ of the sensory data $\varphi$.
In the conventional application of  the FEP the following approximate inequality is usually exercised \cite{Friston:Neural2009,Friston:Frontier2012},
\begin{equation} \label{Math-FEP2}
-\ln p(\varphi) \le {\cal F}.
\end{equation}
However, note that  the inequality, Eq.~(\ref{Math-FEP2}) is not equivalent to Eq.~ (\ref{Math-FEP1}),  in general.
It is only a point approximation piecewise in time.

Here, a difficulty arises because the functional shape of the R-density is not given \textit{a priori}.
It may be fixed after knowing all orders of its moments of the external states, which is not possible.
To circumvent the difficulty, usually one invokes variational Bayes by assuming a Gaussian fixed-form for  the R-density, the `Laplace approximation',
\begin{equation}\label{fixedform}
q(\vartheta) =\frac{1}{\sqrt{2\pi\zeta}} \exp\left\{
-(\vartheta-\mu)^2/(2 \zeta) \right\} \equiv {\cal N}(\vartheta;\mu,\zeta),
\end{equation}
which is fully characterized simply by its means $\mu$ and variances $\zeta$, namely first and second order {\textit{sufficient statistics}, respectively.
Then, by substituting Eq.~(\ref{fixedform}) into Eq.~(\ref{FE-functional}) and after some technical approximations, see \cite{Kim2017} for the details, one can convert the IFE functional ${\cal F}$ into
\begin{equation}\label{FE-function}
{\cal F}[q(\vartheta),p(\vartheta,\varphi)]\rightarrow  -\ln p(\mu,\varphi)  \equiv F(\mu,\varphi).
\end{equation}
At the end of manipulation the outcome becomes a \textit{function} of only the means $\mu$, given the sensory input $\varphi$; where the dependence of variances has been removed by their optimal values.
The resulting IFE \textit{function} $F$ in Eq.~(\ref{FE-function}) is termed the `Laplace-encoded' IFE in which
the parameters $\mu$, specifying the organism's belief or expectation of the environmental states,  are the organism's probabilistic representation of the external world.
In turn, it is argued that the variational parameters $\mu$ are encoded in the brain as biophysical variables.

To proceed with minimization of the IFE in the filtering scheme, a model for noisy-data measurement and also the equations of motion of the states must be supplied.
The FEP assumes a formal homology between the external dynamics and the organism's top-down belief: The former describes, according to physics laws, the equations of motion of environmental states and the sensory-data registering process. And, the latter prescribes the internal dynamics of the representations of the environmental states and the generative model of the sensory data in the organism's brain \cite{Frsiton:BiolCybern2010}.
Following this idea, we hypothesize that the registered data $\varphi$ are predicted by the organism according to a linear or nonlinear morphism,
\begin{equation}\label{map1}
\varphi = g(\mu) + z,
\end{equation}
where $g(\mu)$ is a map from $\mu$ onto $\varphi$ and $z$ is the involved random fluctuation.
Also, the brain's representations $\mu$ of the causes are assumed to obey the stochastic equation of motion,
\begin{equation}\label{Langevin}
\frac{d\mu}{dt} = f(\mu) + w
\end{equation}
where $f(\mu)$ is a linear or nonlinear function of the organism's expectation of environmental dynamics and $w$ is the associated random fluctuation.

Assuming mutually uncorrelated Gaussian fluctuations, $w$ and $z$, of the organism's beliefs, one may furnish the models for the likelihood $p(\varphi|\mu)$ and the empirical prior $p(\mu)$, which jointly enter the Laplace-encoded IFE in Eq.~(\ref{FE-function}) in the factorized form,
\begin{equation}\label{G-density}
p(\varphi,\mu)=p(\varphi|\mu)p(\mu).
\end{equation}
Using the notation introduced in Eq.~(\ref{fixedform}), they are given explicitly as
\begin{eqnarray}
&&p(\varphi|\mu) = {\cal N}(\varphi-g(\mu);0,\sigma_z), \label{likelihood1}\\
&&p(\mu) = {\cal N}(\dot\mu - f(\mu);0,\sigma_w), \label{prior1}
\end{eqnarray}
where we have set $\dot\mu=d\mu/dt$, and the normal densities are assumed to possess zero means with variances $\sigma_z$ and $\sigma_w$, respectively.
When the fluctuations are \textit{statistically stationary}, the variances are handled as constant.; however \textit{nonstationarity} can also be taken into account by assuming an explicit time-dependence in the variances.
Finally, by substituting  Eqs.~(\ref{likelihood1}) and (\ref{prior1}) into Eq.~(\ref{FE-function}), one can convert the Laplace-encoded IFE , up to a constant, into
\begin{equation}\label{var2}
F(\mu,\varphi) = \frac{1}{2}\sigma_z^{-1}\varepsilon_z^2 +
\frac{1}{2}\sigma_w^{-1}\varepsilon_w^2 + \frac{1}{2}\ln\left(\sigma_z
\sigma_w\right),
\end{equation}
where the new variables have been defined as
\[ \varepsilon_z \equiv \varphi - g(\mu) \quad {\rm and}\quad
\varepsilon_w \equiv \dot\mu - f(\mu).
\]
The auxiliary variable $\varepsilon_z$ specifies the discrepancy between the sensory data $\varphi$ and the brain's prediction $g(\mu)$.
Similarly,  $\varepsilon_w$ specifies the discrepancy between the change of the environmental representations $\dot\mu$ and the organism's belief $f(\mu)$.

It is straightforward to extend the formulation to the multiple, correlated noisy inputs.
However, for simplicity, we shall continually work in the single-variable picture and will promote it to the general situation later.

\subsection{Gradient descent scheme of the RD}
\label{GD}
\indent

With the Laplace-encoded IFE as an instrumental tool, the organism's brain searches for the tightest bound for the surprisal, conforming to Eq.~(\ref{Math-FEP2}), by varying its internal states $\mu$.
The critical question is what machinery the brain hires for the minimization procedure.
Typically the gradient descent method in machine learning theory is employed in the conventional approach.

To give an idea of the gradient-descent scheme, here we set up a simple gradient-descent equation, in the usual manner, by regarding the IFE function $F$ as an objective function as
\begin{equation}\label{grad}
\dot \mu = -\kappa\nabla_\mu F.
\end{equation}
In the above $\dot \mu$ implies a \textit{temporal} or \textit{sequential} update of the brain variable $\mu$ and $\nabla_\mu$ is the gradient operator with respect to $\mu$, and $\kappa$ is the learning rate that controls the speed of optimization.
In steady state, defined by $\dot\mu \equiv 0$, the solution $\mu^{(0)}$ to the relaxation equation, Eq.~(\ref{grad}) must satisfy $\nabla_\mu F =0$.
Subsequently, it may be interpreted that such a solution corresponds to an equilibrium (or fixed) point of the IFE function $F$, specifying a local minimum in the IFE landscape.

By inspection, however, we find  that the gradient-descent construct in the preceding way bears an ambiguity when it is applied to dynamic causal models such as Eq.~(\ref{Langevin}).
This is because imposing the condition, $\dot \mu \equiv 0$ on the LHS  of Eq.~(\ref{grad}),  does not guarantee a desired  equilibrium point in the state space spanned by $\mu$.
The reason is that $\dot \mu$ also appears on the RHS of Eq.~(\ref{grad}) via $F$:
The gradient operation on the RHS of Eq.~(\ref{grad}) can be performed explicitly for given $F$, Eq.~(\ref{var2})  to give
\[
\hat\mu\cdot\nabla_\mu F =
- \sigma_z^{-1}(\varphi-g)\frac{\partial g}{\partial \mu} -
\sigma_w^{-1}\left(\dot\mu - f\right)\frac{\partial f}{\partial \mu}.
\]

This subtlety does not appear in the conventional theory which incorporates the \textit{nonstationarity}, i.e., the aspect of continually changing external states, into formulation using the mathematical construct of unbounded, higher-order motion of the \textit{generalized coordinates}.\footnote{The terminology of the generalized coordinates in generalized filtering is dissimilar from its common usage in physics. In classical mechanics,
the generalized coordinates refer the independent coordinate variables which
are required to completely specify the configuration of a system with a holonomic constraint, not including their temporal derivatives.
The number of generalized coordinates determines the degree of freedom in the system \cite{Landau}. Accordingly, the term, `generalized states' seems better suit generalized filtering.}
It is an attempt to allow a more precise specification of a system's dynamical state.
The generalized coordinates are defined to be a row vector in the state space spanned by all orders of time-derivatives of  bare state $\mu$,
\begin{equation}\label{gencoor}
\tilde\mu = (\mu,\mu',\mu'',\cdots) \equiv
(\mu_{[0]},\mu_{[1]},\mu_{[2]},\cdots)
\end{equation}
where vector components are defined,  with understanding $\mu_{[0]}\equiv\mu$,  as
\[
{\mu}_{[n+1]} = {\mu}_{[n]}^\prime\equiv D{\mu}_{[n]}.
\]
Note that the notation $D\mu_{[n]}\equiv\mu_{[n]}^\prime$ has been introduced to denote the \textit{dynamical update} of  the component $\mu_{[n]}$, which is in contrast to the notation $\dot\mu_{[n]}$ for the sequential update.
Also, two components of a vector at different dynamical  orders in the generalized coordinates are mutually independent variables.
Similarly, the sensory-data $\tilde\varphi$ are expressed in the generalized coordinates as a row vector,
\begin{equation}\label{genphi}
\tilde\varphi = (\varphi,\varphi',\varphi'',\cdots)
\equiv(\varphi_{[0]},\varphi_{[1]},\varphi_{[2]},\cdots).
\end{equation}
Each component in the vectors, $\tilde\mu$ and $\tilde\varphi$, is to be considered as a \textit{dynamically-independent} variable.
Also, assuming that the random fluctuations,  $z$ and $w$, are analytic, they have been written in the generalized coordinates as $\tilde z$ and $\tilde w$, respectively.
Then, the generalization of Eqs.~(\ref{map1}) and (\ref{Langevin}) follows after some technical approximations as  (for details, see \cite{Kim2017})
\begin{eqnarray}
&&\tilde \varphi= \tilde g + \tilde z, \label{Geq1}\\
&&D\tilde \mu = \tilde f + \tilde w;\label{Geq2}
\end{eqnarray}
where $D\tilde\mu = (\mu',\mu'',\mu''',\cdots)$.
For reference, we explicitly speel out $n$, Eqs.~(\ref{Geq1}) and (\ref{Geq2}) at dynamical orders as
\begin{eqnarray*}
\varphi_{[n]} &=& \frac{\partial g}{\partial \mu}\mu_{[n]} + z_{[n]},\\
D\mu_{[n]} &=& \frac{\partial f}{\partial \mu}\mu_{[n]} + w_{[n]}.
\end{eqnarray*}
Note that different dynamical orders of the noises $\tilde z$ and $\tilde w$ may be considered to be  statistically correlated in general.
Then,  the Laplace-encoded  IFE can be mathematically constructed from multivariate correlated Gaussian noises with zero means and covariance matrices $\Sigma_w$ and $\Sigma_z$,
\begin{eqnarray}\label{IFE3}
F(\tilde\mu,\tilde\varphi)
&=& \frac{1}{2} \{\dot{\tilde\mu} - {\tilde f}\} \Sigma^{-1}_{w} \{{\dot{\tilde \mu}} - \tilde f\}^T + \frac{1}{2} \ln |\Sigma_{w}| \nonumber\\
&+& \frac{1}{2} \{\tilde\varphi- {\tilde g}\} \Sigma^{-1}_z \{\tilde\varphi- {\tilde g}\}^T + \frac{1}{2} \ln |\Sigma_z|, \label{gmulti1}
\end{eqnarray}
where $\{{\dot{\tilde\mu}} - {\tilde f}\}^T $ is the transpose of row vector $\{{\dot{\tilde\mu}} - {\tilde f}\}$.; $|\Sigma_w|$ and $\Sigma^{-1}_w$ are the determinant and the inverse of the covariance matrix $\Sigma_{w}$, respectively, etc.
In many practical exercises, however, usually, the conditional independence among different dynamical orders is imposed. Consequently, the noise distribution at each dynamical order is assumed to be an uncorrelated Gaussian density about zero means. This simplification corresponds to the Wiener process or Markovian approximation \cite{Jazwinski:1970}. Here, we recall that the generalized states $\tilde \mu$ are the means of the brain's probabilistic model of the dynamical world, the R-density Eq.~(\ref{fixedform}) after rewritten in the generalized coordinates.  Note Eq.~(\ref{IFE3}) is a direct generalization of Eq.~(\ref{var2}).

Furnished with the extra theoretical constructs, the IFE becomes a function of the generalized coordinates $\tilde \mu$, given sensory data $\tilde \varphi$, $F=F(\tilde \mu,\tilde \varphi)$.
Accordingly, the gradient-descent scheme must be extended to incorporate the generalized motions in its formulation.
This is done by the theoretical prescription that the dynamical update $D\tilde \mu$ is distinctive from the sequential update $\dot {\tilde \mu}$.
Consequently, one recasts Eq.~(\ref{grad}) into the form,
\begin{equation}\label{grad3}
\dot {\tilde \mu} - D\tilde \mu = - \kappa\nabla_{\tilde\mu}
F (\tilde\mu,\tilde\varphi).
\end{equation}
With the revised  gradient-descent equation, the conventional FEP claims that the IFE is minimized by reaching a desired fixed point $\tilde\mu^*\equiv\tilde\mu(t\rightarrow \infty)$ in the generalized state space spanned by $\tilde \mu$.
This corresponds to the situation when the two rates of the generalized brain variables, $\dot{\tilde\mu}$  and $D{\tilde\mu}$ become coincident, namely $\dot{\mu}_{[n]} = D{\mu_{[n]}}$ at every dynamic order $n$.
To support the idea it is argued that the purpose of  Eq.~(\ref{grad3}) is to place the gradient descent in the frame of reference that moves with the mean $\tilde\mu$ , see \cite{Friston:GF2010}.
The entire minimization procedure is compactly expressed in the literature as
\[
\tilde \mu^* = \arg \min_{\tilde\mu} F(\tilde\mu,\tilde\varphi|m),
\]
where we have inserted $m$ in $F$ to indicate explicitly that the minimization is conditioned on the generative model of an organism.

In brief, the brain performs the RD of perceptual inference by biophysically implementing Eq.~(\ref{grad3}) in the gray matter.
The steady-state solution $\tilde \mu^*$ specifies minimum value of the IFE, say $F_{\rm min}= F(\tilde\mu^*,\tilde\varphi)$, giving the tightest bound of the surprisal [see Eq.~(\ref{Math-FEP2})] associated with a given sensory experience $\tilde\varphi$.
Despite its frequent employment in practicing the FEP, we have disclosed some subtleties involved in the conventional formulation of the gradient descent scheme, which has motivated our reformulation.

\section{The informational action principle}
\label{IAP}
\noindent

The RD condensed in Sec.~\ref{GD} is based on the mathematical statement Eq.~(\ref{Math-FEP2}) of the FEP, which is a point approximation of Eq.~(\ref{Math-FEP1}).
Here we reformulate the RD by complying with the full mathematical statement of FEP given in Eq.~(\ref{Math-FEP1}).
Accordingly, we need a formalism that allows minimization of the time-integral of the IFE, not at each point in time.
We have come to assimilate that the theoretical `action' in the principle of least action neatly serves the goal \cite{Landau}.
This formalism allows us to eschew introduction of the generalized coordinates of a dynamical state comprising an infinite number of time-derivatives of the brain state $\mu$.
Consequently, not required is the distinctive classification of time-derivative of the parametric update ($\dot\mu$) and the dynamical update ($D\mu$) of the state variable.
In what follows, we shall consistently use the dot symbol to denote time-derivative of a dynamical variable.

\subsection{Lagrangian formalism}
\label{LagFor1}
\noindent

To formulate the RD from the principle of least action, the `Lagrangian' of the system must be supplied.
We define the informational Lagrangian (IL) of the brain, denoted by ${\cal L}$, as the Laplace-encoded IFE function,
\[
{\cal L}(\mu,\dot\mu;\varphi) \equiv F(\mu,\dot\mu;\varphi),
\]
where we have placed the semicolon in ${\cal L}$ to indicate that $\mu$ and $\dot\mu$ are the two brain's dynamical variables, given sensory input $\varphi$.
The sensory inputs are stochastic and time-dependent, in general; $\varphi=\varphi(t)$, reflecting the changing external states, of which generative processes are to be supplied by physics laws.
The proposed IL is not a physical quantity but an information-theoretic object.
When we take Eq.~(\ref{var2}) as an explicit expression for $F$, the IL is written up as
\begin{equation}\label{Lagrangian}
{\cal L}(\mu,\dot\mu;\varphi) = \frac{1}{2}\sigma_w^{-1}(\dot\mu-f(\mu))^2 + \frac{1}{2}\sigma_z^{-1}(\varphi-g(\mu))^2 .
\end{equation}
Note that we have dropped out the term, $\frac{1}{2}\ln\left(\sigma_z \sigma_w\right)$ in writing Eq.~(\ref{Lagrangian}) by assuming it as a constant, which then does not affect the dynamics of $\mu$ and $\dot\mu$.
This assumption of statistical nonstationarity may be lifted by introducing time-dependence in the variances,
\[\sigma_w=\sigma_w(t)\quad{\rm and}\quad\sigma_z=\sigma_z(t).\]
Still, however,  the dropped-out term does not affect the dynamics because a term that can be expressed as a total time-derivative in the Lagrangian will not do anything \cite{Landau}.

Next, we postulate that the perceptual dynamics of the neural particles conforms to the principle of least action \cite{Landau}.
Accordingly, we suppose that the brain's perceptual operation corresponds to searching for an optimal dynamical path that minimizes the informational action (IA), denoted by $\cal S$,
\begin{equation}\label{Action}
{\cal S} \equiv \int_{t_i}^{t_f} dt~ {\cal L}(\mu,\dot\mu;\varphi);
\end{equation}
where $t_f-t_i\equiv T$ is  the temporal horizon over which a living organism encounters an environmental event.
When functional derivative of ${\cal S}$ is carried out with respect to
$\mu$ and $\dot\mu$, it gives
\[ \delta {\cal S} = \left[\frac{\partial {\cal L}}{\partial \mu}\delta
\mu\right]_{t_i}^{t_f} - \int_{t_i}^{t_f} dt \left( \frac{d}{dt}\frac
{\partial {\cal L}}{\partial \dot\mu}- \frac{\partial {\cal L}}
{\partial \mu}\right)\delta\mu. \]
By imposing $\delta S \equiv 0$ under the condition that initial and final
states are fixed,
\[\delta \mu(t_i)=0=\delta \mu(t_f),
\]
we derive the Lagrangian equation as
\begin{equation}\label{LagEq}
\frac{d}{dt}\frac{\partial{\cal L}}{\partial\dot\mu} -
\frac{\partial{\cal L}}{\partial \mu} = 0.
\end{equation}
Using the specified Lagrangian, Eq.~(\ref{Lagrangian}),  in Eq.~(\ref{LagEq}), we obtain a Newtonian equation of motion for the brain variable $\mu$,
\begin{equation}\label{EqMotion}
\sigma_w^{-1}\dot v =  \bar\Lambda_1+\bar\Lambda_2,
\end{equation}
where we have defined the kinematic velocity to be
\[v\equiv \dot \mu\] and the additional notations on the RHS as
\begin{equation}\label{netforce}
\bar\Lambda_1 \equiv \sigma_w^{-1}f\frac{\partial f}{\partial \mu} \quad{\rm and}\quad
\bar\Lambda_2 \equiv  - \sigma_z^{-1}(\varphi-g)\frac{\partial g}{\partial\mu} .
\end{equation}
Equation~(\ref{EqMotion}) entails the RD of the brain in the Lagrangian formulation.
We interpret that the inverse of the variance $\sigma_w^{-1}$ plays, as a metaphor, a role of  \textit{inertial mass} of the neural particles.
Accordingly, the LHS of Eq.~(\ref{EqMotion}) represents an \textit{inertial force}, i.e. the product of  `inertial mass' and `acceleration', $\ddot\mu$.
Note that the inverse of variance is interpreted as \textit{precision} in the Friston formulation \cite{Kim2017}, which gives a measure for the accuracy of the brain's expectation or prediction of sensory data.
So, the precision is the `informational mass' of the neural particle metaphorically.
Also, the terms $\bar\Lambda_i$, $i=1,2$,  on the RHS are interpreted as the `forces' that drive the internal $\bar\Lambda_1$ as well as sensory $\bar\Lambda_2$ excitations in the brain.
The acceleration can be evaluated from $\ddot\mu=\sum \bar\Lambda_i/\sigma_w^{-1}$ when the net force is known.

While the organism's brain integrates the RD for incoming sensory input, an optimal trajectory $\mu^*(t)$ is continuously achieved in the neural-configuration space.
And, the steady-state condition in the long-time limit $t\rightarrow \infty$  is given by
\begin{equation}\label{resting}
\dot \mu^*=v^* = {\rm const},
\end{equation}
where the net force vanishes.
Note that equation~(\ref{resting}) defining  an attractor, $\mu_{eq}=\mu^*(\infty)$, is more general than the simple guess, $\dot\mu^*=0$.
The optimal trajectory $\mu^*(t)$ minimizes the IA, which, in turn, provides the organism with the tightest estimate of the sensory uncertainty, see equation~(\ref{Math-FEP1}).

\subsection{Hamiltonian formalism}
\label{HamFor1}
\noindent

The mechanical formulation can be made more modish in terms of Hamiltonian language
which admits position and momentum as independent brain variables, instead of position and velocity in the Lagrangian formulation.
The positions and the momenta span \textit{phase space} of a physical system, which defines the neural state space of the organism's brain.

The `canonical' momentum $p$, which is conjugate to the position $\mu$, is defined via Lagrangian ${\cal L}$ as \cite{Landau}
\begin{equation}\label{momentum}
p \equiv \frac{\partial {\cal L}}{\partial \dot\mu} =
\sigma_w^{-1}\left(\dot\mu-f\right),
\end{equation}
which evidently differs from the `kinematic' momentum $\sigma_w^{-1}v=\sigma_w^{-1}\dot\mu$.
Then, the informational `Hamiltonian' ${\cal H}$ may be constructed from the Lagrangian using Legendre's
transformation \cite{Landau},
\begin{equation}\label{Legendre}
{\cal H}(\mu,p;\varphi) = \sum \frac{\partial {\cal L}}{\partial \dot\mu}\dot\mu - {\cal L}
(\mu,\dot\mu;\varphi).
 \end{equation}
The first term on the RHS of Eq.~(\ref{Legendre}) can be further manipulated to give
\[ \sum \frac{\partial {\cal L}}{\partial \dot\mu}\dot\mu = \sigma_w^{-1}\dot\mu^2 -\sigma_w^{-1}\dot\mu f.\]
By plugging the outcome and also the Lagrangian ${\cal L}$ given in Eq.~(\ref{Lagrangian}) into Eq.~(\ref{Legendre}), we obtain the Hamiltonian as a function of $\mu$ and $p$ as desired, given $\varphi$,
\begin{equation}\label{Hamiltonian}
{\cal H}(\mu,p;\varphi)  =  {\cal T}(p) +{\cal  V}(\mu,p;\varphi),
\end{equation}
where Eq.~(\ref{momentum}) has been used to replace $\dot\mu$ with $p$.
The first term on the RHS of Eq.~(\ref{Hamiltonian}) is the `kinetic energy' which depends only on momentum,
\begin{equation} \label{kinetic}
{\cal T}(p) = \frac{p ^2}{2\sigma_w^{-1}} .
\end{equation}
Also, the second term on the RHS of Eq.~(\ref{Hamiltonian}) is the `potential energy' which depends on both position and momentum,
\begin{equation} \label{effpotential}
{\cal V}(\mu,p;\varphi) = V(\mu;\varphi) + p f(\mu) ,
\end{equation}
where we have defined the momentum-independent  term separately as $V$,
\begin{equation}
\label{potential}
V(\mu;\varphi) = -\frac{1}{2}\sigma_z^{-1}(\varphi-g)^2.
\end{equation}
We remark that the sensory stimuli $\varphi$ enter the Hamiltonian only through the potential-energy part $V$ which becomes `conservative' when $\varphi$ is static.
Here, we shall assume that the variances associated with the noisy data are constant.
For time-varying sensory inputs, in general, the Hamiltonian is \textit{nonautonomous}.
In Fig.~\ref{Fig-potential} we depict the conservative potential energy, using three-term approximations for the generative function,
\[
g(\mu)\approx b_1+b_2\mu+b_2\mu^2.
\]
For convenience, we have assumed a fixed sensory input, $\varphi=15$, and set parameters as $(b_1,b_2,b_3)=(0,1,0.01)$.
We have observed numerically that the static sensory signal $\varphi$ changes the distance between two unstable fixed points, but do not affect the location of the stable equilibrium point.
Also, the depth of the stable equilibrium valley gets deeper as the magnitude of $\varphi$ increases.
\begin{figure}[h!]
\begin{center}
\includegraphics{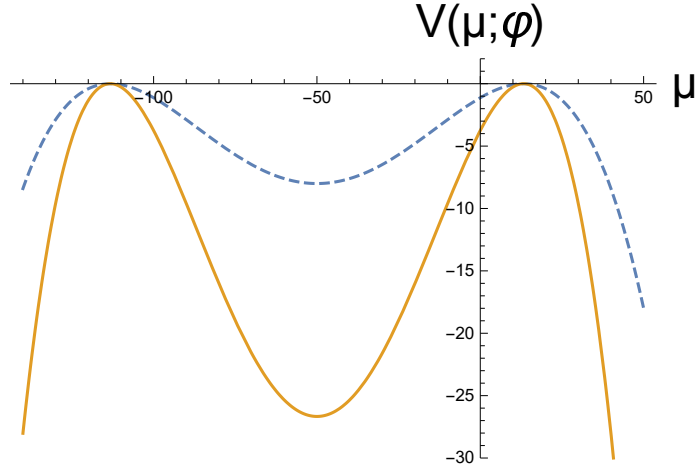}
\caption{The potential energy, given in Eq.~(\ref{potential}), in arbitrary units; where the dashed and solid curves are for the variance $\sigma_z=100$ and $30$, respectively. Both cases exhibit a stable minimum in the central well and two unstable maxima on the side hills, which contribute to determining the FE landscape.}
\label{Fig-potential}
\end{center}
\end{figure}

Next, we take the total derivative of the Hamiltonian given in Eq.~(\ref{Legendre}) with respect to $\mu$ and $\dot \mu$ to get
\begin{eqnarray*}
d{\cal H}(\mu,p;\varphi) &=& \sum d\left(p\dot\mu\right) - d{\cal L} (\mu,\dot\mu;\varphi) \nonumber\\
&=& \dot \mu dp +p d\dot \mu - \left(\frac{\partial {\cal L}}{\partial \mu}d\mu + \frac{\partial {\cal L}}{\partial \dot\mu}d\dot\mu\right) \nonumber \\
&=& - {\dot p}_\mu d\mu + {\dot \mu} dp.
\end{eqnarray*}
By comparing the last expression with the formal expansion,
\[d{\cal H}=\frac{\partial {\cal H}}{\partial \mu}d\mu +  \frac{\partial {\cal H}}{\partial p}dp, \]
we identify the Hamilton equations of motion for independent variables $\mu$ and $p$ of a neural particle,
\begin{eqnarray}
\dot\mu &= & \frac{\partial {\cal H}}{\partial p} \label{preHam1}\\
\dot p &= &-\frac{\partial {\cal H}}{\partial \mu} \label{preHam2}.
\end{eqnarray}
For given ${\cal H}$  in Eq.~(\ref{Hamiltonian}), we spell out the RHS of Eq.~(\ref{preHam1}) to get
\begin{equation}\label{Ham1}
\dot \mu = \frac{1}{\sigma_w^{-1}} p + f
\end{equation}
which is identical to Eq.~(\ref{momentum}).
Similarly, the second equation, Eq.~(\ref{preHam2}) is spelled out
\begin{equation} \label{Ham2}
\dot p = - \frac{\partial V}{\partial \mu} - \frac{\partial f}{\partial \mu}p.
\end{equation}
The first term on the RHS of Eq.~(\ref{Ham2}) specifies the conservative force,
\[
- \frac{\partial V}{\partial \mu} \rightarrow -\sigma_z^{-1}(\varphi-g)\frac{\partial g}{\partial
\mu}.
\]
Whereas, the second term on the RHS of Eq.~(\ref{Ham2}) specifies the dissipative force, where $\partial f/\partial \mu$ plays the role of damping coefficient.

The derived set of coupled equations for the variables $\mu$ and $p$ furnish the RD of the brain in phase space spanned by $\mu$ and $p$, which involve only first order time-derivatives.
When time-derivative is taken once more for both sides of Eq.~(\ref{Ham1}) with followed substitution of Eq.~(\ref{Ham2}) for $\dot p$, the outcome becomes identical to the Lagrangian equation of motion, Eq.~(\ref{EqMotion}).
This observation confirms that two mechanical formulations, one from the Lagrangian and the other from the Hamiltonian, are in fact equivalent.

In the Hamiltonian formulation, the brain's fulfilling of the RD is equivalent to finding an optimal trajectory $(\mu^*(t),p^*(t))$ in phase space.
For a static sensory input, the dynamics governed by Eqs.~(\ref{Ham1}) and (\ref{Ham2}) is autonomous, and for the time-dependent sensory input it becomes non-autonomous.
The RD can be integrated by providing appropriate models for the generative functions $f$ and $g$.
The \textit{attractor} $(\mu^*(\infty),p^*(\infty))$  would be a \textit{focus} or \textit{center}  in phase space, which
can be calculated by simultaneously imposing the conditions on LHSs of Eqs.~(\ref{Ham1}) and (\ref{Ham2}) ,
\begin{equation}
\label{fixedpts}
\dot \mu^*=0\quad{\rm and}\quad \dot p^*=0.
\end{equation}
One can readily confirm that these fixed-point conditions match with the Newtonian equilibrium condition, $\sum_i \bar\Lambda_i=0$ in the Lagrangian formulation, see section~\ref{LagFor1}.
The situation corresponds to the brain's resting state at a local minimum on the energy landscape defined by the Hamiltonian function.

\subsection{Multivariate formulation}
Having established the Hamiltonian dynamics for a single brain variable $\mu$, we now extend our formulation to the general case of the multivariate brain.
We denote $\{\mu\}$ as a row vector of $N$ brain states as done in Sec.~\ref{IFE},
\[
\{\mu\} = (\mu_1,\mu_2,\cdots,\mu_N) ,
\]
that respond to the multiple of sensory inputs in a general way,
\[
\{\varphi\} = (\varphi_1,\varphi_2,\cdots,\varphi_N).
\]
For simplicity, we neglect the statistical correlation of the fluctuations associated with environmental variables and also with sensory inputs.
Then, within the independent-particle approximation of uncorrelated brain variables,  the Laplace-encoded IFE Eq.~(\ref{gmulti1}) furnishes the multivariate Lagrangian,
\begin{equation}\label{Lagrangian2}
{\cal L}(\{\mu\},\{\dot\mu\};\{\varphi\}) = \frac{1}{2} \sum_{\alpha=1}^N \left[ \sigma_{w\alpha}^{-1}\left(\dot \mu_\alpha - f_\alpha(\{\mu\})\right)^2
+ \sigma_{z\alpha}^{-1}\left(\varphi_\alpha - g_\alpha(\{\mu\})\right)^2 \right],
\end{equation}
where we have dropped out the terms which contain only the variances, $\sigma_{z\alpha}$ and $\sigma_{w\alpha}$, assuming that the noises are statistically nonstationary.
One may extend Eq.~(\ref{Lagrangian2}) to interacting neural nodes in terms of covariance matrix formulation \cite{Kim2017}, which is not our concern here, either.
Subsequently, the conjugate momentum to the generalized coordinate $\mu_\alpha$ is determined by an explicit evaluation of
\begin{equation}\label{momentum2}
p_\alpha = \frac{\partial {\cal L}}{\partial \dot \mu_\alpha} =
\sigma_{w\alpha}^{-1}\left(\dot\mu_\alpha-f_\alpha\right).
\end{equation}
Note the momentum $p_\alpha$ gives a measure of the discrepancy, weighted by the inverse variance $\sigma_{w\alpha}$, between the change of the probabilistic representation of the environment $\dot\mu_\alpha$ and the organism's belief of it $ f_\alpha$. The weighting factor $\sigma_{w\alpha}^{-1}$ is called the \textit{precision} in the FEP.
In turn, the Hamiltonian of the multivariate brain can be constructed from Eq.~(\ref{Hamiltonian}) as
\begin{equation}\label{Hamiltonian1}
{\cal H}(\{\mu\},\{p\};\{\varphi\}) = {\cal T}(\{\mu\},\{p\};\{\varphi\}) + {\cal V}(\{\mu\},\{p\};\{\varphi\})
\end{equation}
where first term on the RHS is the kinetic energy,
\begin{equation}\label{effkin2}
{\cal T}(\{p\};\{\varphi\})\equiv \sum_\alpha\frac{{p_\alpha}^2 }{2\sigma_{w\alpha}^{-1}}
\end{equation}
and the potential energy ${\cal V}$ is identified as
\begin{equation}\label{effpot2}
{\cal V}(\{\mu\},\{p\};\{\varphi\}) \equiv \sum_\alpha\left[ - \frac{1}{2}\sigma_{z\alpha}^{-1}(\varphi_\alpha-g_\alpha)^2 + p_\alpha f_\alpha\right].
\end{equation}
Then, it is straightforward to derive the RD of the variables $\mu_\alpha$ and $p_\alpha$, given sensory data $\varphi_\alpha$, as
\begin{equation}\label{Hameq1}
\dot \mu_\alpha = \frac{\partial {\cal H}}{\partial p_\alpha} = \frac{1}{\sigma_{w\alpha}^{-1}} p_\alpha + f_\alpha,
\end{equation}
and for their conjugate momenta,
\begin{equation} \label{Hameq2}
\dot p_\alpha = -\frac{\partial {\cal H}}{\partial \mu_\alpha} =  - \frac{\partial g_\alpha}{\partial \mu_\alpha}p_{\varphi\alpha} - \frac{\partial f_\alpha}{\partial \mu_\alpha}p_\alpha.
\end{equation}
In writing equation~(\ref{Hameq2}), for notational convenience we have introduced an auxiliary quantity $p_{\varphi\alpha}$,
\[p_{\varphi\alpha} \equiv \sigma_{z\alpha}^{-1} (\varphi_\alpha-g_\alpha).\]

Equations~(\ref{Hameq1}) and (\ref{Hameq2}) are a coupled set of equations for the computational units,  $\mu_\alpha$ and $p_\alpha$, describing the brain states and their conjugate momenta, respectively, given the sensory discrepancy $p_{\varphi\alpha}$ between the observed data $\varphi_\alpha$ and their predictions $g_\alpha(\mu_\alpha)$.
With some working models for $f_\alpha$ and $g_\alpha$, they shape the RD in the brain's multi-dimensional phase space in the Hamiltonian prescription.

In Fig.~\ref{Fig-SimpleCircuitry} we present a schematic illustration of the perceptual circuitry implied by the RD at a neural node.
The classification of excitatory and inhibitory activation of the computational units is not absolute because the overall sign depends on the generative function $f_\alpha$ and map $g_\alpha$, which are not specified.
\begin{figure}[h!]
\begin{center}
\includegraphics{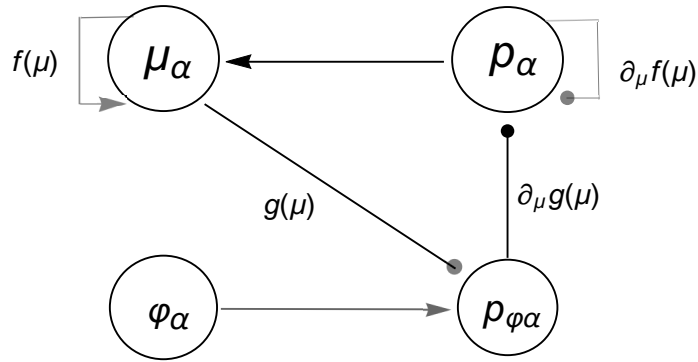}
\caption{The perceptual circuitry at neural node $\alpha$ in which sensory data $\varphi_\alpha$ stream; where it is depicted that the computational units $\mu_\alpha$ and $p_\alpha$ are positively activated by arrows and negatively by lines ended with filled dots. The conjugate momenta $p_\alpha$, defined in Eq.~(\ref{momentum2}), to the brain variables $\mu_\alpha$ mimic the precision-weighted prediction errors in the language of predictive coding \cite{RaoBallard:1999}.}
\label{Fig-SimpleCircuitry}
\end{center}
\end{figure}
It is admissible to assume that the brain is, at the outset, in a resting state.
As the sensory inputs $\varphi_\alpha$ come in, the organism's brain performs the RD online, by integrating Eqs.~(\ref{Hameq1}) and (\ref{Hameq2}), to attain an optimal trajectory in neural phase space,
\[ \mu_\alpha=\mu_\alpha^*(t)\quad{\rm and}\quad p_\alpha=p_\alpha^*(t),\]
which minimize the IA, see Eq.~(\ref{Action}).
The entire minimization procedure may be stated abstractly as
\begin{equation}\label{rFEP}
(\mu_\alpha^*,p_\alpha^*) = \arg \min_{\mu_\alpha,p_\alpha} {\cal S}(\mu_\alpha,p_\alpha;\varphi|m),
\end{equation}
where ${\cal S}$ is the IA and $m$ has been inserted to indicate explicitly that minimization is conditioned on the organism as a model of the world.

Note that in our revised RD is involved not only the organism's prediction of the environmental change via its representation $\mu_\alpha$ but also the dynamics of its prediction-error $p_\alpha$.

\section{Biophysical implementation}
\label{Biophysics}
\indent

We know that the anatomy and entire functions of an organism's brain develop from single cells. In order to provide empirical Bayes in the FEP with a solid biophysical basis, we must start with known biophysical substrates and then introduce probabilities to describe a neuron, neurons, and a network.  Until now, however, most work has taken the reverse direction: Theory prescribes first a conjectural model and then tries to allocate possible neural correlates.
At present, our knowledge remains limited on how biophysical mechanisms of neurons implicate predictions and model aspects about the environment, while a `neurocentric' approach to the inference problem seems suggestive to bridge the gap \cite{Fiorillo2008,Fiorillo:2014}.

Here, we regard coarse-grained Hodgkin-Huxley (H-H) neurons as the generic, basic building-blocks of encoding and transmitting a perceptual message in the brain.
The famous H-H model continues to be used to this day in computational neuroscience studies of neuronal dynamics \cite{HH1952,Hille}.
In extracellular electrical recordings, the local field potential and multi-unit activity result in as combined signals from a population of neurons \cite{Einevoll}.
Such averaged neuronal variables must subserve the perceptual states and conduct the cognitive computation in the brain.
We shall call them `small' neural particles and envisage that a small neural particle enacts a \textit{node} that collectively forms the whole \textit{neural network} on a large scale.
Before proceeding, we shall mention that there are many biophysical efforts to describe such averaged neuronal properties; for instance, the neural mass models and neural field theories are a few examples  \cite{Jansen:1993,Jirsa:1996,Robinson:1997,David:2003,Deco:2008}.
Also, we note the bottom-up effort of trying to understand the large-scale brain function at the cortical microcircuit level based on the averaged, spikes and synaptic inputs over a coarse-grained time interval \cite{Diesmann:2014,Steyn-Ross:2016}.

\subsection{Single cell description}
\label{singlenode}

We first present how our formulation may be implemented at a single-cell level by hypothesizing that each neuron reflects the fundamentals of the perceptual computation of the whole system.
A typical neuron receives current information about its surroundings from the sensory periphery via glutamate, which  excites or inhibits the membrane potential $V$ with regulating the gating variables $\gamma_l$ and ionic concentrations $n_l$; where $l$ is the ion channel index.
We assume that $(V,\{n_l\},\{\gamma_l\})$ specify the neural states of a neuron as a neural observer in the neural configurational space \cite{Fiorillo:2014}.
We encapsulate the neural states as components in a multi-dimensional row vector,
\[\{\mu\} = (V,\{n_l\},\{\gamma_l\}) = (\mu_1,\mu_2,\mu_3,\cdots).\]

The H-H equation for excitation of the membrane voltage $V$ in a spatially homogeneous cell is given by
\begin{equation}\label{Hodgkin1}
C\frac{dV}{dt} = \sum_l\gamma_lG_l(E_l-V) + I_{ex}(t)
\end{equation}
where $C$ is the membrane capacitance, $G_l$ is the maximal conductance of ion channel $l$, $\gamma_l$ is the probability factor associated with opening or closing channel $l$ which in general a product of activation and inactivation gating variables, and $I_{ex}$ is the external driving current.
For simplicity, contributions from leakage current  as well as synaptic input are assumed to be included in the external currents.
The reverse potential $E_l$ of $l$-th ion channel is given, allowing its time-dependence via nonequilibrium ion-concentrations, in general, as
\begin{equation}\label{Nernst}
E_l(t) = \frac{k_BT}{q_l}\ln \frac{n_{li}(t)}{n_{lo}(t)},
\end{equation}
where $k_B$ is the Boltzmann constant, $T$ is the metabolic temperature of an organism, $q_l$ is the ionic charge of channel $l$, and  $n_{li}(t)$ and $n_{lo}(t)$ are the instantaneous ion concentrations inside and outside the membrane, respectively.
In the steady state without external current, $I_{ex}=0$, $V$ tends to the resting (Nernst) potential $V(t\rightarrow \infty)$ with retaining ionic concentrations in electro-chemical equilibrium.
The gating variable $\gamma_l$ of ion channels is assumed to obey the kinetics, a different model of that may be preferable,
\begin{equation}\label{kinetics}
\frac{d\gamma_l}{dt}=-\frac{1}{\tau_l}(\gamma_l-\gamma_{leq}) + \eta_l,
\end{equation}
where the relaxation time $\tau_l$ and steady-state gating variable $\gamma_{leq}$ depend on the membrane potential, in general,
\[
\tau_l=\tau_l(V) \quad{\rm and}\quad \gamma_{leq}=\gamma_{leq}(V),
\]
and $\eta_l$ is the noise involved in the process.

For ionic concentration dynamics, we suppose that ion concentrations $\{n_l\}$ vary slowly compared to the membrane potential and gating-channel kinetics, and consequently treat them statically in our work.
This restriction can be lifted when a more detailed description is required for ion concentration dynamics.
Accordingly, the reverse potentials $E_l$ are also treated statically in below.

Then, the state equations for the multivariate neural vaiable $\{\mu\}$ neatly map onto the standard form  suggested  in the FEP,
\begin{equation}\label{Hudgikin2}
\frac{d\mu_\alpha}{dt} = f_\alpha(V,\{\gamma_l\},\{n_l\}) + w_\alpha(t),
\end{equation}
 where $\alpha$ runs $1,2,\cdots$ with implying $\mu_1=V$, $\mu_2=\gamma_1$, and  $\mu_3=\gamma_2$, etc.
 The driving functions $f_\alpha$, that are specified to be
\begin{eqnarray}
f_V(V,\{\gamma_l\};\{n_l\}) &=& \frac{1}{C}\sum_l\gamma_lG_l(E_l-V)  + \frac{1}{C}I_{ex} , \label{neuralGV} \\
f_{\gamma_l}(V,\{\gamma_l\};\{n_l\}) &=& -\frac{1}{\tau_l}(\gamma_l-\gamma_{leq}). \label{neuralGgam}
\end{eqnarray}
The terms $w_\alpha$ in Eq.~(\ref {Hudgikin2}) describe the noisy synaptic and/or leakage current $w_V$ flowing into the neural cell, not the deterministic contribution $I_{ex}$ which is included in $f_V$, and the noise $w_{\gamma_l}= \eta_l$ associated with the activation and inactivation of ion channels, respectively.
For both noises, we assume the Gaussian distributions ${\cal N}(\dot\mu_\alpha - f_\alpha;0,\sigma_{w\alpha})$
with variances $\sigma_{w\alpha}$ about zero means.

Regarding neuronal response to the sensory stimulus $\varphi_\alpha$, we adopt the usual generative map in the FEP [see Eq.~(\ref{map1})] as
\begin{equation}\label{sensory}
\varphi_\alpha = g_\alpha(V,\{\gamma_l\},\{n_l\}) + z_\alpha,
\end{equation}
where  $g_\alpha$ is the generative map that is unknown but must be supplied for practical application and $z_\alpha$ characterizes the stochastic nature of the sensory reading which we assume the normal distribution  ${\cal N}(\varphi_\alpha -g_\alpha;0,\sigma_{z\alpha})$.
With the present model, we admit that the neural observer responds to the sensory data instantly by means of the neuronal states.
Currently, we do not possess a firm ground on biophysical processes of the sensory prediction.

As a working example, here we consider a H-H neuron which allows fast relaxation, i.e. $\tau_l\ll 1$,  of gating variables to their steady-states, $\gamma_l(t)\rightarrow\gamma_l(\infty)=\gamma_{leq}(V) $.
In this case our neural particle is fully characterized by a single dynamical variable of $V$.
Note the time-dependence of the gating variables occurs only implicitly through the long-time membrane voltages in Eq.~(\ref{neuralGV}).
Then, the RD of our neural particle is fulfilled in a two-dimensional state space spanned by $\{\mu\}=(V,p_V)\equiv (\mu, p)$, prescribed by the Hamiltonian function, equation~(\ref{Hamiltonian}),
\begin{equation*}
{\cal H}(\mu,p)=\frac{p^2}{2\sigma_w^{-1}}  - \frac{1}{2}\sigma_{z}^{-1}(\varphi - g)^2 + p f.
\end{equation*}
While the `dissipative' function $f$ is explicitly given in the H-H model as
\begin{equation}\label{instgV}
f(\mu) = \frac{1}{C}\sum_l\gamma_{leq}(\mu)G_l(E_l-\mu)+I_{ex}/C,
\end{equation}
the `conservative' function $g$ must be additionally supplied.
Also, one needs to make the voltage-dependence of $\gamma_{leq}$ available in practice.
Note the Hamiltonian is \textit{nonautonomous}, in general, because it explicitly depends on time through both the sensory input $\varphi(t)$ and the driving current $I_{ex}$ in $f$, and also through $\sigma_w(t), \  \sigma_z(t)$ when the noisy data are statistically nonstationary.

\begin{figure}[h!]
\begin{center}
\includegraphics{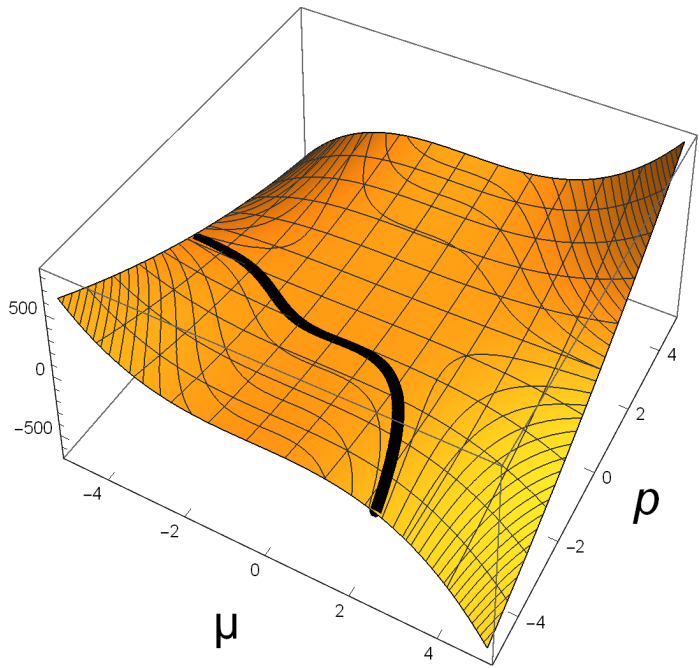}
\caption{Hamiltonian function ${\cal H}(\mu,p)$ in arbitrary units for the chosen set of parameters given in the main text; where the black curve on the energy landscape is the trajectory which is calculated by solving the Hamilton equations of motion for an initial condition at $(\mu,p)=(2.5,-5.0)$. [For interpretation of the references to color in this figure, the reader is referred to the web version of this article.]}
\label{Fig-Hamiltonian}
\end{center}
\end{figure}
In Fig.~\ref{Fig-Hamiltonian} we present the energy landscape described by the Hamiltonian function, assuming static sensory data, constant driving currents, and statistical stationarity.
Since our knowledge is limited to the functional form of $g(\mu)$ and $f(\mu)$, we have taken the algebraic approximations \cite{Wilson:1999},
\begin{equation*}\label{approx-gmap}
g(\mu) \approx a_{0} + a_{1}\mu + a_{2}\mu^2,
\end{equation*}
\begin{equation*}\label{approx-gate}
f(\mu) \approx b_{0 } + b_{1}\mu + b_{2}\mu^2 + b_3 \mu^3.
\end{equation*}
For numerical purposes, we have specified $(a_0,a_1,a_2)=(0,1,1)$ and $(b_0,b_1,b_2,b_3)=(0,0.1,1,1)$, and also assumed a fixed sensory input with equal masses (precisions ) on the brain's internal model and belief of sensory prediction as
\[\varphi=1.0 \quad {\rm and}\quad \sigma_w^{-1}=0.1=\sigma_z^{-1}.\]

The Hamilton equations of motion, Eqs.~(\ref{Hameq1}) and (\ref{Hameq2}), bring about the nonlinear RD as
\begin{eqnarray}
&&\dot \mu = \Lambda_1(\mu,p;t), \label{snrecog1}\\
&&\dot p = \Lambda_2(\mu,p;t), \label{snrecog2}
\end{eqnarray}
where the `force' functions $\Lambda_1$ and $\Lambda_2$  are specified as
\begin{eqnarray}
&&\Lambda_1 = f(\mu) + \frac{1}{\sigma_w^{-1}}p , \label{isocline1}\\
&&\Lambda_2 = -\sigma_z^{-1}(\varphi-g)\frac{\partial g}{\partial\mu} - \frac{\partial f}{\partial \mu}p. \label{isocline2}
\end{eqnarray}
We have chosen an initial state and solved the equations of motion, for the same parameters used in Fig.~\ref{Fig-Hamiltonian}, to obtain a trajectory in phase space.
The outcome is drawn on the energy surface in Fig.~\ref{Fig-Hamiltonian}.
According to the present model, the neural observer performs the RD, given the sensory input $\varphi$ and, consequently, obtains the optimal trajectory $(\mu^*,p^*)$ conforming to Eq.~(\ref{rFEP}).
In the long-time limit the brain will reach a fixed (equilibrium) point $(\mu^*_{eq},p^*_{eq})$ in the state space, that is specified by intersections of two \textit{isoclines},
\[
\Lambda_i(\mu,p;\infty)=0, \quad i=1,2.
\]
We have determined the fixed points numerically. It turns out that there exist three real solutions for the specified system parameters, $(-1.23,0.05)$, $(-0.50,-0.01)$, and $(0.07,-0.04)$, which are depicted as the blue dots in Fig.~\ref{Fig-Center}.
By further analysis, we have found that only the middle point is a stable equilibrium solution and the other two specify saddle points.
Figure~\ref{Fig-Center} shows a flow of trajectories, obtained from arbitrary initial points on the red-colored circle of radius $\mu^2+p^2=1.6$, in phase space.

\begin{figure}[h!]
\begin{center}
\includegraphics{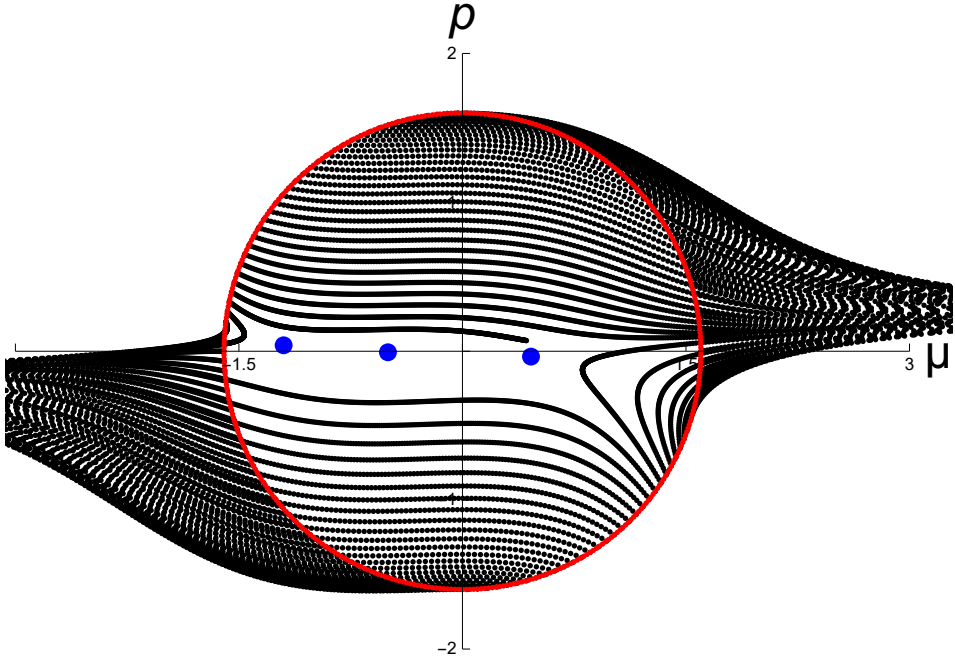}
\caption{Optimal trajectories in phase space which are obtained by integrating  the RD, equations~(\ref{snrecog1}) and (\ref{snrecog1}), from the initial conditions  arbitrarily chosen on the  red-colored circle; where the blue dots are the equilibrium points among which only the middle dot at $(-0.50,-0.01)$ is a stable fixed point, and other two points are saddle points. The stable fixed point turns out to be a center, which we have confirmed by linear stability analysis and also numerically. [For interpretation of the references to color in this figure, the reader is referred to the web version of this article.]}
\label{Fig-Center}
\end{center}
\end{figure}
To gain an insight into how the system approaches to a steady state, we inspect the optimal trajectories near an equilibrium point,
\[
\mu^*\approx\mu^*_{eq}+\delta\mu^* \quad {\rm and} \quad p^*\approx p^*_{eq}+\delta p^*.
\]
We expand  Eqs.~(\ref{snrecog1}) and (\ref{snrecog2}) to the linear order in the deviations $\delta\mu^*$ and $\delta p^*$ and, after rearrangement, obtain the normal form,
\begin{equation}\label{singleRecog}
\frac{d}{dt}
\left(\begin{array}{c}
\delta \mu^*  \\
\delta p^*
\end{array}\right)
+
\left(\begin{array}{cc}
{\cal R}_{11} & {\cal R}_{12}  \\
{\cal R}_{21} & {\cal R}_{22}
\end{array}\right)
\left(\begin{array}{c}
\delta\mu^*\\
\delta p^*
\end{array}\right)
=0.
\end{equation}
In Eq.~(\ref{singleRecog}) the elements of the relaxation (Jacobian) matrix  ${\cal R}$ are specified to be
\begin{eqnarray*}
&&{\cal R}_{11}=-\left[\frac{\partial f}{\partial \mu}\right]_{eq}, \quad  {\cal R}_{12}=-\frac{1}{\sigma_w^{-1}}  \\
&&{\cal R}_{21}=\sigma_z^{-1}\left[-\left(\frac{\partial g}{\partial \mu}\right)^2 + (\varphi-g)\frac{\partial^2 g}{\partial \mu^2} - \frac{\partial^2 f}{\partial \mu^2}p \right]_{eq},\\
&&{\cal R}_{22} = \left[\frac{\partial f}{\partial \mu}\right]_{eq};
\end{eqnarray*}
where the partial derivatives are to be evaluated at the equilibrium points.
Here, for notational convention we denote the column vector as
\[
\delta\psi \equiv
\left(\begin{array}{c}
\delta \mu^*  \\
\delta p^*
\end{array}\right).
\]
Then, the formal solution to Eq.~(\ref{singleRecog}) is written as
\[
\delta\psi(t)=e^{-{\cal R}t}\delta\psi(0).
\]
One may expand the initial state $\psi(0)$ in terms of the eigenvectors of ${\cal R}$ as
\[ \delta\psi(0) = \sum c_\alpha \phi_\alpha,\]
where the eigenvalues $\lambda_\alpha$ and eigenvectors $\phi_\alpha$ are determined by the secular equation,
\[ {\cal R}\phi_\alpha=\lambda_\alpha\phi_\alpha. \]
Consequently, the solutions to the linear RD at a single node level is completed as
\begin{equation}\label{singleSol}
\delta\psi(t)=\sum_{\alpha=1}^2 c_\alpha e^{-\lambda_\alpha t}\phi_\alpha,
\end{equation}
where the expansion coefficients $c_\alpha$ are fixed by the initial condition.

In the linear regime, a geometrical interpretation of the equilibrium solutions is possible by inspecting the eigenvalues of the Jacobian matrix ${\cal R}$.
Considering that the matrix ${\cal R}$ is not symmetric, we anticipate that the eigenvalues are not real.
Furthermore, because the trace of the relaxation matrix equals zero, the sum of the two eigenvalues must be zero.
Thus, when the determinant of ${\cal R}$ is positive, the two eigenvalues $\lambda_1$ and $\lambda_2$ would be pure imaginary with opposite sign.
Consequently, in the present particular model, the resulting equilibrium point is likely to be a \textit{center}.
We have confirmed numerically that the eigenvalues of the Jacobian corresponding to the stable equilibrium point in figure~\ref{Fig-Center} meet the condition for a center.

\subsection{The hierarchical neural network}
\label{neuralnetwork}
\noindent

Here, we suppose that there are a finite number of \textit{levels} in the perceptual hierarchy of the whole system and that for simplicity each level is characterized efficiently as a single neural node.
Further, we assume that the neural node at hierarchical level $i$ is described by the coarse-grained, activation and connection variables, denoted as $V^{(i)}$ and $S^{(i)}$, respectively.
The activation variable describes action potential at a node, and the connection variable describes inter-level synaptic input and output variables.
Both variables are derived from a population of neurons and thus vary on a coarse-grained space and time scale.
The technical details of how one may derive such a coarse-graining description are not our scope,  for a reference see \cite{Deco:2008}.
They form the coordinates in brain's configurational space,
\[{\cal \mu}^{(i)}=(V^{(i)},S^{(i)}), \]
where the superscript runs $i=1,2,\cdots,M$, with $M$ denoting the highest level.

We assume that the activation variables $V^{(i)}$ obey the effective dynamics with noise $w^{(i)}$ within each hierarchical level $i$,
\begin{equation}\label{lateral-dynamics}
\frac{dV^{(i)}}{dt} = f^{(i)}(V^{(i)},S^{(i)}) + w^{(i)},
\end{equation}
which is a direct generalization of Eq.~(\ref{Hudgikin2}) with incorporating the hierarchical dependence via $S^{(i)}$.
For inter-level dynamics,  we propose that the connection variables are updated by one-level higher connetion as well as activation variables, subjected to the stochastic equations,
\begin{equation}\label{hierch-dynamics}
\frac{dS^{(i)}}{dt} = g^{(i+1)}(V^{(i+1)},S^{(i+1)}) + z^{(i)},
\end{equation}
where $z^{(i)}$ represents the noise associated with the process.
The brain's top-down prediction functions  $f^{(i)}$  and $g^{(i)}$ must be supplied in practical implementation.
Note there is only spontaneous fluctuation at the top cortical level, $i=M$, accordingly
\begin{equation}\label{sensory3}
\quad g^{(M+1)}= 0.
\end{equation}
Also, we constrain that the sensory data $\varphi$ enter the interface (or boundary between) of the brain and the environment specified as the lowest hierarchicl level, $i=1$.
Subsequently we assume that the brain's prediction of the sensory inputs is performed by way of an instantaneous mapping,
\begin{equation}\label{sensory4}
S^{(0)} = g^{(1)}(V^{(1)},S^{(1)}) + z^{(0)},
\end{equation}
where, for notational convenience, we have set
\[
S^{(0)}\equiv \varphi(t).
\]

We remark that the hierarchical equations Eq.~(\ref{hierch-dynamics}) we propose is dissimilar to the conventional formulation which assumes the static model in the entire hierarchy like the one Eq.~(\ref{sensory4}) at the sensory interface, see \cite{Kim2017}.
We treat here the connection variables dynamically not statically to treat lateral and hierarchical dynamics symmetrically.
The rates of the activation and connection variables may be subjected to different time-scales, that can be incorporated, for instance, by introducing distinctive relaxation-times in their generative functions.
It turns out that our equations suit the formalism of the Hamilton action principle neatly.

Having specified our hierarchical model, we write the informational Lagrangian for the constructed neural network by generalizing Eq.~(\ref{Lagrangian2}) with a single sensory input for now, as
\begin{equation}\label{Lagrangian3}
{\cal L}(V,\dot V;S,\dot S;\varphi) = \frac{1}{2}\sum_{i=1}^M {m^{(i)}_w}\left(
\varepsilon^{(i)}_{w} \right)^2
+
\frac{1}{2}\sum_{i=0}^{M} {m^{(i)}_z}\left(\varepsilon^{(i)}_{z} \right)^2,
\end{equation}
where $m^{(i)}_w$ and $m^{(i)}_z$ are the inertia masses, associated with the Gaussian noises, $w^{(i)}$ and $z^{(i)}$, respectively, defined to be
\begin{equation}\label{infmasses}
m^{(i)}_w \equiv 1/{\sigma_w^{(i)}} \quad {\rm and}\quad m^{(i)}_z \equiv 1/{\sigma_z^{(i)}}.
\end{equation}
The auxiliary variables in the Lagrangian are defined to be ($i\ge 1$)
\begin{eqnarray}
&&\varepsilon^{(i)}_w \equiv \dot V^{(i)} - f^{(i)}\left(V^{(i)},S^{(i)}\right), \label{prederror1}\\
&&\varepsilon^{(i)}_z  \equiv \dot S^{(i)}-g^{(i+1)}\left(V^{(i+1)},S^{(i+1)}\right). \label{prederror2}
\end{eqnarray}
We interpret that $\varepsilon^{(i)}_w$ specifies the discrepancy between the change in the
present lateral state and the brain's on-level prediction, which may be considered as the lateral prediction-error.
On the other hand, $\varepsilon^{(i)}_z$ measures the prediction error between the change in the present hierarchical state and its prediction from one higher level via the generative map $g$, which may be viewed as the hierarchical prediction-error.
Note $\varepsilon^{(0)}_z$ in the second term on the RHS of Eq.~(\ref{Lagrangian3}) is defined separately as
\[
\varepsilon^{(0)}_z  \equiv S^{(0)}-g^{(1)}\left(V^{(1)},S^{(1)}\right),
\]
which specifies an error estimation in sensory prediction at the lowest hierarchical level.

The generalized momenta, conjugate to $V^{(i)}$ and $S^{(i)}$, are readily calculated  for $i\ge 1$, respectively, as
\begin{eqnarray}
&&p_V^{(i)} \equiv \frac{\partial {\cal L}}{\partial \dot V^{(i)}} = m^{(i)}_w \varepsilon^{(i)}_w, \label{momentum3}\\
&&p_S^{(i)} \equiv \frac{\partial {\cal L}}{\partial \dot S^{(i)}} = m^{(i)}_z \varepsilon^{(i)}_z, \label{momentum4}
\end{eqnarray}
Note that the inverse variances $m^{(i)}_w$, $m^{(i)}_z$  have been termed the informational masses, see discussion below equation~(\ref{netforce}).
The role of the inertial masses is to modulate the discrepancy between the change of the perceptual states and their prediction.
Thus, in our theory, momentum $p_V^{(i)}$ is a measure of lateral prediction-error modulated by inertial mass $m_w^{(i)}$, and momentum $p_S^{(i)}$ is a measure of hierarchical prediction-error modulated by inertial mass $m_z^{(i)}$.
And, the heavier the mass is, the bigger the precision becomes.

Given the Lagrangian Eq.~(\ref{Lagrangian3}), we can formulate the informational Hamiltonian by performing a Legendre transformation,
\[
{\cal H}= \sum_i \left(\dot V^{(i)}p_V^{(i)}+\dot S^{(i)}p_S^{(i)}\right) - {\cal L}.
 \]
After some manipulation, we obtain the outcome as
\begin{equation}\label{Hamilton3}
{\cal H}(V,p_V;S,p_S;\varphi) = \sum_{i=1}^{M}\left({\cal T}^{(i)} + {\cal V}^{(i)}\right),
\end{equation}
where the infromational kinetic energy is defined to be  ($i\ge 1$)
\begin{equation}\label{effkin3}
{\cal T}^{(i)}(p_V,p_S) = \frac{1}{2 m^{(i)}_w} \left(p^{(i)}_V\right)^2 + \frac{1}{2 m^{(i)}_z} \left(p^{(i)}_S\right) ^2,
\end{equation}
and the potential energy to be ($i\ge 2$)
\begin{equation}\label{effpot3}
{\cal V}^{(i)}(V,p_V;S,p_S;\varphi) \equiv  p_V^{(i)}f^{(i)} +  p_S^{(i)}g^{(i+1)}.
\end{equation}
Note the potential energy at the lowest level is specified separately to be
\begin{equation}\label{effpot4}
 {\cal V}^{(1)} = p_V^{(1)}f^{(1)} - \frac{1}{2 m^{(0)}_z} \left(p^{(0)}_S\right) ^2;
 \end{equation}
where, for notational convention, we have written the weighted prediction-error associated with the sensory measurement as
\[p_S^{(0)} \equiv m^{(0)}_z \varepsilon^{(0)}_z=m^{(0)}_z(\varphi- g^{(1)}), \]
which, unlike $p^{(i)}_S$ for $i\ge 1$,  is not a canonical momentum.
Consequently,  the multi-level Hamiltonian Eq.~(\ref{Hamilton3}) has been prescribed via the perceptual states  in the hierarchical chain, $i=1,2,\cdots, M$,  denoted as  a four dimensional column vector $\psi^{(i)}$ at each level,
\[ \psi^{(i)}=(V^{(i)},p^{(i)}_V,S^{(i)},p^{(i)}_S)^{\rm T}\equiv  (\psi^{(i)}_1, \psi^{(i)}_2, \psi^{(i)}_3, \psi^{(i)}_4)^{\rm T} ,\]
where ${\rm T}$ means the transverse operation.

Next, it is straightforward to generate the Hamiltonian equations of motion for the brain's perceptual states $\psi^{(i)}$.
The results are the coupled differential equations for the four computational components at each level ($i\ge 1$), which are, in turn, hierarchically-connected among adjacent levels:
\begin{eqnarray}
&&\dot V^{(i)} = \frac{\partial {\cal H}}{\partial p^{(i)}_V} = \frac{1}{m^{(i)}_w } p^{(i)}_V + f^{(i)}, \label{HPMi1} \\
&&\dot S^{(i)} = \frac{\partial {\cal H}}{\partial p^{(i)}_S} = \frac{1}{m^{(i)}_z} p^{(i)}_S + g^{(i+1)}, \label{HPMi3} \\
&&\dot p^{(i)}_V = -\frac{\partial {\cal H}}{\partial V^{(i)}} =  -\frac{\partial f^{(i)}}{\partial V^{(i)}}p^{(i)}_V -\frac{\partial g^{(i)}}{\partial V^{(i)}} p^{(i-1)}_S; \label{HPMi2} \\
&&\dot p^{(i)}_S = -\frac{\partial {\cal H}}{\partial S^{(i)}} = -\frac{\partial f^{(i)}}{\partial S^{(i)}} p^{(i)}_V-\frac{\partial g^{(i)}}{\partial S^{(i)}} p^{(i-1)}_S. \label{HPMi4}
\end{eqnarray}

According to the derived RD,  the sensory inputs $\varphi$ enter the brain-environment interface at the level $j=1$, which are instantly predicted by the organism's lowest-level generative model  $g^{(1)}(V^{(1)},S^{(1)})$.
Subsequently, the resulting prediction-error $p^{(0)}_S$  acts as a source to update the prediction-errors, $p^{(1)}_V$ and $p^{(1)}_S$.
The change of on-level perceptual states $V^{(1)}$ and $S^{(1)}$ are predicted by  the generative models $f^{(1)}$ and $g^{(2)}$ with additional modulations from the perceptual momenta  $p^{(1)}_V$ and $p^{(1)}_S$, being the lateral and hierarchical prediction-errors, respectively.
At higher levels $i\ge 2$, the intra-level dynamics of the activation state  $V^{(i)}$ is updated,  through Eq.~(\ref{HPMi1}), by the on-level generative function $f^{(i)}$ and prediction error $p^{(i)}_V$, while the change of the current hierarchical state $S^{(i)}$ is determined, through Eq.~(\ref{HPMi3}), by the inter-level prediction  $g^{(i+1)}$ and the on-level prediction error $p^{(i)}_S$.
The organism's top-down message flow is mediated by the connection state $S^{(i)}$ via Eq.~(\ref{HPMi3}) as $(S^{(i+1)},V^{(i+1)})\rightarrow S^{(i)}$.
And, Eqs.~(\ref{HPMi2}), and (\ref{HPMi4}) govern the coupled, bottom-up propagation  of  the prediction errors, mediated by $p^{(i)}_S$, $p_S^{(i)}\rightarrow (p_S^{(i+1)},p_V^{(i+1)})$.
In Fig. \ref{Fig-FullDiagram} we draw a diagram that schematically illustrates the perceptual architecture of the hierarchical network, at lowest two levels, implied by Eqs.~(\ref{HPMi1})-(\ref{HPMi4}).
\begin{figure}[h!]
\begin{center}
\includegraphics{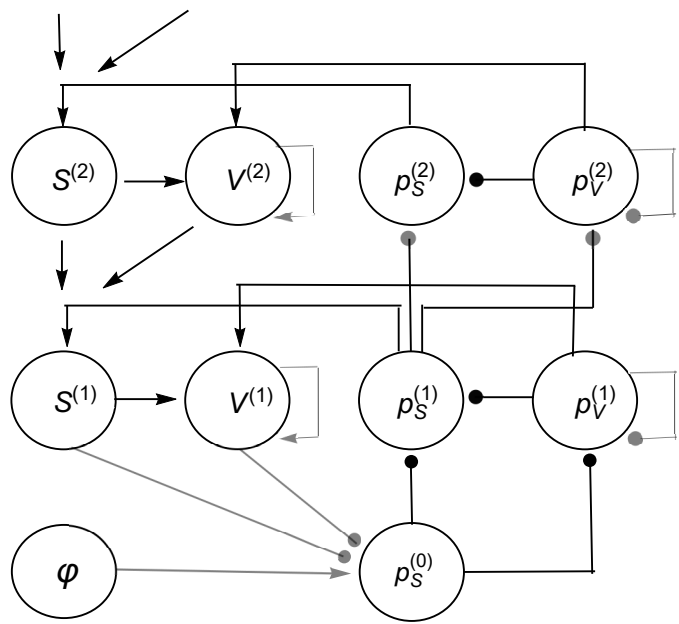}
\caption{A schematic of the neural circuitry which conducts the RD Eqs.~(\ref{HPMi1})-(\ref{HPMi4}) in the hierarchical network of the brain; where the computational units $(S^{(i)},V^{(i)},p^{(i)}_S,p^{(i)}_V)$, $i=1,2,\cdots,M$, are connected by arrows for excitatory (positive) inputs and by lines ended with filled dots for inhibitory (negative) inputs. Note that the prediction error $p_S^{(0)}$ of incoming sensory data $\varphi$, at the lowest level, induces an inhibitory change in the perceptual momenta $(p_S^{(1)},p_V^{(1)})$. Subsequently, the prediction error propagates up in the hierarchy, $p_S^{(1)}\rightarrow (p_S^{(2)},p_V^{(2)})$, etc. On the other hand, the top-down message passing is mediated by means of the connection states $S^{(i)}$.  For instance, the connection state $S^{(1)}$ is top-down predicted by both units $(S^{(2)},V^{(2)})$ from one higher-level. }
\label{Fig-FullDiagram}
\end{center}
\end{figure}

Here, we emphasize that the dynamics of precision-weighted prediction errors, encapsulated in canonical momenta in which mass takes over the role of precision, are taken into account in our Hamiltonian formulation on an equal footing with the dynamics of prediction of the state variables.
This aspect is also in contrast to the conventional minimization algorithm which entails differential equations only for the update of the brain states without carrying parallel ones for the prediction errors.
Consequently, the message passing in our model shows different features in the details compared with the neural circuitry from the conventional RD \cite{Bastos:2012}.
However, the general message flow, in terms of the computational units, of feedforward, feedback, and lateral connections hold the same in the hierarchical brain network.
We recognize an attempt to incorporate the brain's computation of prediction errors in the FEP can be found in a recent tutorial model \cite{Bogacz2017}.

Here, for mathematical compactness, we rewrite the filtering algorithm, Eqs.~(\ref{HPMi1})-(\ref{HPMi4}), as
\begin{equation}\label{KimModel}
\frac{d\psi_\alpha^{(i)}}{dt}=\Lambda_\alpha^{(i)}(\{\psi_\alpha^{(i)}\}),
\end{equation}
where the hierarchical index $i$ runs from $1$ to $M$, $\alpha$ runs from $1$ to $4$, and the force function $\Lambda_\alpha^{(i)}$ is the corresponding RHS to each vector component $\psi_\alpha^{(i)}$ at  cortical level $i$.
The obtained hierarchical equations are the highlight of our theory, prescribing the RD of the brain's sensory inference under the FEP framework.

To apply our formulation to an empirical brain, one needs to supply the generating function of lateral dynamics  $f^{(i)}$  and the hierarchical connecting function  $g^{(i)}$, that enter the force functions $\Lambda_\alpha^{(i)}$ in the perceptual mechanics, Eq.~(\ref{KimModel}).
For the generating function we once again use the H-H model Eq.~(\ref{neuralGV})  to write
\begin{equation} \label{instgV1}
f^{(i)}(V^{(i)},S^{(i)}) = \sum_{l}\gamma_{leq}{\tilde G}_l\left(E_l-V^{(i)}\right) + {\tilde G}_{S}S^{(i)}\left( E_{S}-V^{(i)} \right),
\end{equation}
where ${\tilde G}_l$ are the channel conductances normalized by the capacitance $C$.
And, the second term on the RHS accounts for other deterministic driving sources such as leakage and/or lateral synaptic currents with ${\tilde G}_S$ being the normalized synaptic conductance.
The hierarchical connection function, for which we have limited biophysical knowledge, shall be taken in a simple form here as
 \begin{equation}\label{gmap1}
g^{(i)}(V^{(i)},S^{(i)}) = \Gamma(V^{(i)})S^{(i)},
\end{equation}
where the function $\Gamma$ specify the voltage-dependent synaptic plasticity from hierarchical level $i$ to level $i-1$.
In addition, as in the single-node case, one must supply approximate models for voltage-dependence of the gating variables $\gamma_{leq}$ and the connection strength $\Gamma$.
For instance, one may take the quadratic approximations \cite{Wilson:1999},
\begin{eqnarray*}
&& \gamma_{leq}(V^{(i)}) \approx b_{l0} + b_{l1}V^{(i)} + b_{l2}V^{(i)}V^{(i)},\\
&& \Gamma(V^{(i)}) \approx a_{0} + a_{1}V^{(i)} + a_{2}V^{(i)}V^{(i)}.
\end{eqnarray*}

Having laid down the lateral and hierarchical generative models, the organism's brain can now perform the RD given a streaming of noisy inputs.
While conducting the filtering, an optimal trajectory is obtained in multi-dimensional phase space,
\[ \psi_\alpha^{*(i)} = \psi_\alpha^{*(i)}(t),\]
which, in the end, tends to a fixed point, $\psi_{\alpha, eq}^{(i)}=\psi_\alpha^{*(i)}(t\rightarrow\infty)$.
The necessary equilibrium condition to Eq.~(\ref{KimModel}) is specified by
\begin{equation}
\Lambda_\alpha^{(i)}(\{\psi_\alpha^{(i)}\}) = 0.
\end{equation}

Although the full time-dependent solutions must be invoked numerically, one may inspect the perceptual trajectories near a fixed point by linear analysis.
To this end, we consider a small deviation  of $\alpha^{th}$ component of the perceptual state vector $\psi_\alpha^{*(i)}$ at the cortical level $i$, $\delta \psi_\alpha^{(i)}$, from the fixed point $\psi_{\alpha,eq}^{(i)}$,
\[
\psi_\alpha^{*(i)}\approx \psi_{\alpha0}^{(i)} + \delta \psi_\alpha^{(i)}.
\]
Then, we expand Eq.~(\ref{KimModel}) about the fixed point to linear order in the small deviation, and after some manipulation we get the hierarchical equations for $\delta \psi_\alpha^{(i)}$,
\begin{equation}\label{NetworkRecog}
\frac{d\delta\psi_\alpha^{(i)}}{dt} +\sum_{\beta=1}^4 {\cal R}_{\alpha\beta}^{(i)}\delta\psi_\beta^{(i)} = \sum_{\beta=1}^4 \sum_{j\neq i}^M\Phi_{\alpha\beta}^{(ij)}\delta\psi_\beta^{(j)};
\end{equation}
where the $\alpha\beta$ component of the $4\times 4$ Jacobian matrix at cortical level $i$ is specified by
\[
 {\cal R}_{\alpha\beta}^{(i)} = \left[  \frac{\partial \Lambda_\alpha^{(i)}}{\partial \psi_\beta^{(i)}} \right]_{eq},
 \]
and the inter-level connection between level $i$ and level $j$ in the hierarchical pathway is specified by
\[
\Phi_{\alpha\beta}^{(ij)} = \left[ \frac{\partial \Lambda_\alpha^{(i)}}{\partial \psi_\beta^{(j)}} \right]_{eq};
\]
where the subscript $eq$ indicates that the matrix elements are to be evaluated at the equilibrium points.
To cast  the inhomogeneous term into a more suggestive form we further inspect it in detail within the models specified:
We observe first that the matrix elements $\Phi_{\alpha\beta}^{(ij)}$ do not vanish only for $\alpha=3$ because only the force function $\Lambda_3^{(i)}$ possesses $\psi_\beta^{(j)}$ for $j\neq i$ as variables via $g^{(i+1)}$ [see Eq.~(\ref{HPMi3})].
Second, because $g^{(i+1)}$ depends solely on the hierarchical level-index $i+1$, only matrix elements with the hierarchical index $j=i+1$ survives.
Combining these two observations, the source term on the RHS of Eq.~(\ref{NetworkRecog}) is converted into
a vector at level $i+1$ with only a single nonvanishing $\alpha=3$ component,
\[
\sum_{\beta=1}^4 \sum_{j\neq i}^M\Phi_{\alpha\beta}^{(ij)}\delta\psi_\beta^{(j)} \equiv \delta\zeta_\alpha^{(i+1)}
\]
which to be complete we spell out explicitly as
\begin{equation}\label{Source}
\delta\zeta_\alpha^{(i+1)} = \delta_{\alpha 3}\left\{ \left[\frac{\partial g^{(i+1)}}{\partial \psi_1^{(i+1)}}\right]_{eq}\delta\psi_1^{(i+1)} + \left[\frac{\partial g^{(i+1)}}{\partial \psi_3^{(i+1)}}\right]_{eq}\delta\psi_3^{(i+1)} \right\},
\end{equation}
where $\delta_{\alpha 3}$ is the Kronecker delta.

Finally, we shall present a formal solution to the linearized perceptual mechanics Eq.~(\ref{NetworkRecog}), that can be obtained by a direct integration with respect to time.
The result takes the form
\begin{equation}\label{Hformal1}
\delta\psi^{(i)}(t) = e^{-{\cal R}^{(i)}t}\delta\psi^{(i)}(0) + \int_0^t dt^\prime  e^{-{\cal R}^{(i)}(t-t^\prime)}\delta\zeta^{(i+1)}(t^\prime).
\end{equation}
We next solve the eigenvalue problem at each hierarchical level, which is defined to be
\begin{equation}\label{Heigenprob}
{\cal R}^{(i)}\phi_\alpha^{(i)}=\lambda_\alpha^{(i)}\phi_\alpha^{(i)},
\end{equation}
where $\lambda_\alpha^{(i)}$ and  $\phi_\alpha^{(i)}$ are the eigenvalues and corresponding eigenvectors at level $i$, respectively.
Then, we expand the initial state $\delta\psi^{(i)}(0)$  in terms of the complete eigenvectors:
\begin{equation}\label{Hinitial}
\delta\psi^{(i)}(0) = \sum a_\alpha^{(i)} \phi_\alpha^{(i)}.
\end{equation}
Similarly, we may expand the inhomogeneous vector $\delta\zeta^{(i+1)}$ as
\begin{equation}\label{Hsource}
\delta\zeta^{(i+1)}(t^\prime) = \sum b_\alpha^{(i+1)}(t^\prime)\phi_\alpha^{(i)},
\end{equation}
where note the expansion coefficients  $b_\alpha^{(i+1)}$ are time-dependent.
By substituting the expansions Eqs.~(\ref{Hinitial}) and (\ref{Hsource}) into Eq.~(\ref{Hformal1}) we obtain the desired formal solution near equilibrium points,
\begin{equation}
\delta\psi^{(i)}(t) = \sum_{\alpha=1}^4 a_\alpha e^{-\lambda_\alpha^{(i)}t}\phi_\alpha^{(i)} + \sum_{\alpha=1}^4 \phi_\alpha^{(i)}\int_0^t dt^\prime  e^{-\lambda_\alpha^{(i)}(t-t^\prime)}b_\alpha^{(i+1)}(t^\prime).
\end{equation}
The geometrical approach to a fixed point is again determined by the eigenvalues $\lambda_\alpha^{(i)}$; however, the details are driven by the time-dependent generative sources $b_\alpha^{(i+1)}(t)$ from one-level higher in the hierarchy.

To sum, responding to sensory streaming $\varphi=S^{(0)}$, at the lowest hierarchical level ($i=1$), the brain in an initial resting state performs the hierarchical RD by integrating Eq.~(\ref{KimModel}) to infer the external causes.
The ensuing brain's computation corresponds to minimizing the IA, which is an upper bound of the  sensory uncertainty, whose mathematical statement, equation~(\ref{Math-FEP1}), is repeated compactly as
\[H[p(\varphi)] ~\le ~ {\cal S}[F;\varphi],\]
where the sensory uncertainty $H$ was defined in Eq.~(\ref{sensory-uncertainty}) and the IA on the RSH  is expressed here in terms of the hierarchical states as ${\cal S}[F;\varphi]=\int dt F(\{\psi_\alpha^{(i)}\};\varphi)$.
Conforming to the FEP, the minimum value of IA specifies the tightest bound of the sensory uncertainty over a relevant biological time-scale, which preserves the organism's current model of the environment.

\section{Discussion}
\label{Conclusion}
\noindent

We have recast the FEP following the principles of mechanics,  which articulates that all living organisms are evolutionally self-organized to tend to minimize the sensory uncertainty about uninhabitable environmental encounters.
The sensory uncertainty is an average of the surprisal over the sensory density registered on the brain-environment interface, the sensory surprisal being the self-information of the sensory probability density.
The FEP suggests that the organisms implement the minimization by calling forth the IFE in the brain.
The time-integral of the IFE gives an estimate of the upper bound of the sensory uncertainty.
We have enunciated that the minimization of the IFE must continually take place over a finite temporal horizon of an organism's unfolding environmental event.
Our scheme is a generalization of the conventional theory which approximates minimization of the IFE at each point in time when it performs the gradient descent.
The sensory uncertainty is an information-theoretical Shannon entropy \cite{Shannon:1948}; however, in this work, we have circumvented the term, `sensory entropy'  to call the sensory uncertainty.
The reason is that `minimization of the sensory entropy'  is reminiscent of Erwin Schr\"odinger's word, `negative entropy' which carries a disputable connotation in implying how the living organism avoids decay.
He subsequently suggested FE instead as a more appropriate notion in the context \cite{Erwin}.
Conforming to the second law of thermodynamics, the organism's adaptive fitness of minimizing the sensory uncertainty must contribute to increasing the total entropy of the brain and its environment.

We have adopted the Laplace-encoded IFE as an informational Lagrangian in implementing the FEP under the principle of least action.
And, by subscribing to the standard Newtonian dynamics, we have considered the IFE a function of position and velocity as the metaphors for the organism's brain variable and their first-order time derivative, respectively, in the continuous-time picture.
The brain variable maps onto the first-order sufficient statistics of the recognition density launched in the organism's brain to perform the RD, Bayesian filtering of the noisy sensory data.
In the following Hamiltonian formulation, the RD prescribes momentum, conjugate to a position, as a mechanical measure of prediction error weighted by mass, the precision.
The theoretical construct of generalized coordinates introduced in the prevailing theory to specify the extended states of higher-orders of motion has been eschewed in our formulation.
The features of changing world enter our theory via the sensory inputs in continuous time, and the statistical nonstationarity of the noise is embedded in time-dependence of the variances of the Gaussian fluctuations in the organism's belief of the changing state and sensory generation.
The temporal correlations of the dynamical states may be incorporated as time-dependent covariances, but not explored in this work.
Also, in our theory all the parameters in the RD are specified in the Hamiltonian; thus, there require no extra parameters like learning rates in the gradient descent scheme to control the speed of convergence to a steady state.
According to our formulation, the brain's Helmholtzian perception corresponds to finding an optimal trajectory in the hierarchical functional network by minimizing the IA.
When the brain completes the RD by reaching a desired fixed-point or a limit cycle, it remains resting, i.e., being spontaneous, until another sensory stimulus will come in.

We have applied our formalism to a biophysically grounded model for hierarchical neural dynamics by suggesting that the large-scale architecture of the brain be an emergent coarse-grained description of the interacting single neurons.
We have admitted the asymmetric top-down rationale of sensory inference in our formalism, which is an essential facet of the FEP:
The sensory inputs at the interface, the lowest hierarchical level, were assumed to be instantaneously mapped onto the organism's belief, encoded in the brain variables, about environmental causes with associated noise.
Differently, however, from the instant model at the lowest level, we have generalized that the inter-level filtering in the brain's functional hierarchy obeys the stochastic dynamics, supplied with the organism's dynamical generative model of environmental states.
The resulting RD is deterministic and hierarchical, which notably incorporates dynamics of both predictions and prediction errors of the perceptual states on an equal footing.
Consequently, the details of the ensuing neural circuitry from our formulation differs from the one supported by the gradient-descent scheme which generates only dynamics of the causal and hidden states, not their prediction errors. However, the general structure of message passing, namely descending predictions and ascending prediction-errors in the hierarchical network, shows the same.
Also, our obtained RD tenably underpins the causality: For a specified set of the perceptual positions and corresponding momenta, at the outset, responding to sensory inputs that may be slow or fast time-dependent, the RD can be integrated online.
The arbitrariness involved in deciding the number of generalized coordinates for a complete description and also the ambiguity in specifying unknowable initial conditions can be averted.

In short, it is still a long way to understanding how the Bayesian FEP in neurosciences may be made congruous with the biophysical reality of the brain.
It is far from clear how the organism embodies the generative model of the environment in the physical brain.
Our theory only delivers a hybrid of the biologically plausible information-theoretic framework of the FEP and the mechanical formulation of the RD under the principle of least action.
To borrow what Hopfield puts in words, ``it lies somewhere between a model of neurobiology and a metaphor for how the brain computes'' \cite{Hopfield1999}.
We hope that our effort will guide a step forward to unraveling the challenging problem.


\section*{References}

\end{document}